\documentclass[journal,draftcls,onecolumn,12pt,twoside]{IEEEtranTCOM}
\normalsize

\ifCLASSINFOpdf

\else

\fi

\usepackage[cmex10]{amsmath}

\usepackage[T1]{fontenc}% optional T1 font encoding
\usepackage{mathtools}

\DeclarePairedDelimiterX\Basics[1](){ #1}

\usepackage{amsmath,amssymb}

\ifCLASSINFOpdf
\usepackage[pdftex]{graphicx}

\else

\fi

\usepackage{amsmath}
\usepackage{stackengine}
\def\delequal{\mathrel{\ensurestackMath{\stackon[1pt]{=}{\scriptstyle\Delta}}}}
\usepackage{bm}
\usepackage{array}
\usepackage{tabu}
\usepackage{hyperref} %for linking the cross references
\usepackage{graphics} %for dealing with figures.
\usepackage{algorithmic} %for describing algorithms
\usepackage{bm}
\usepackage{graphicx}
\usepackage{subfigure}
\usepackage{subfloat}
\usepackage{amsmath}
\usepackage{caption}
\usepackage{amsthm,amssymb,amsmath}
\usepackage{lmodern}
\usepackage{mathrsfs}
\DeclareMathAlphabet{\mathpzc}{OT1}{pzc}{m}{it}
\usepackage{mathtools}

\usepackage{stackengine}
\def\delequal{\mathrel{\ensurestackMath{\stackon[1pt]{=}{\scriptstyle\Delta}}}}
\usepackage{amsmath}
\usepackage{cite}
\usepackage{graphicx}
\usepackage[font={small}]{caption}
\usepackage{ptext}
\usepackage{algorithm}
\usepackage{algorithmic}
\usepackage{float}

\usepackage{amsmath}
\usepackage{graphicx}

\begin{document}

\title{Analysis and Code Design for the Binary CEO Problem under Logarithmic Loss}

\author{Mahdi~Nangir,
        Reza~Asvadi,~\IEEEmembership{Member,~IEEE,}
        ~Mahmoud~Ahmadian-Attari,~\IEEEmembership{Member,~IEEE,}
        and Jun~Chen,~\IEEEmembership{Senior Member,~IEEE}
\thanks{M. Nangir and M. Ahmadian-Attari are with the Faculty of Electrical Engineering, K.N. Toosi University of Technology, Tehran, Iran. (e-mails: mahdinangir@ee.kntu.ac.ir; mahmoud@eetd.kntu.ac.ir).}
\thanks{R. Asvadi is with the Faculty of Electrical Engineering, Shahid Beheshti University, Tehran, Iran. (e-mail: r\_asvadi@sbu.ac.ir).}
\thanks{J. Chen is with the Department of Electrical and Computer Engineering, McMaster University, Hamilton, ON, Canada (email: junchen@mail.ece.mcmaster.ca).}}
%\markboth{IEEE Transactions on Communications}
%{Submitted paper}
\maketitle
%\vspace{-20mm}
\begin{abstract}
%\boldmath
In this paper, we propose an efficient coding scheme for the binary Chief Executive Officer (CEO) problem under logarithmic loss criterion. Courtade and Weissman obtained the exact rate-distortion bound for a two-link binary CEO problem under this criterion.  We find the optimal test-channel model and its parameters for the encoder of each link by using the given bound. Furthermore, an efficient encoding scheme based on compound LDGM-LDPC codes is presented to achieve the theoretical rates. In the proposed encoding scheme, a binary quantizer using LDGM codes and a syndrome-decoding employing LDPC codes are applied. An iterative decoding is also presented as a fusion center to reconstruct the observation bits. The proposed decoder consists of a sum-product algorithm with a side information from other decoder and a soft estimator. The output of the CEO decoder is the probability of source bits conditional to the received sequences of both links. This method outperforms the majority-based estimation of the source bits utilized in the prior studies of the binary CEO problem. Our numerical examples verify a close performance of the proposed coding scheme to the theoretical bound in several cases.
\end{abstract}
\begin{IEEEkeywords}
Binary CEO problem, logarithmic loss (log-loss), test-channel model, compound LDGM-LDPC codes, soft CEO decoder.
\end{IEEEkeywords}
\IEEEpeerreviewmaketitle

\section{Introduction}

\IEEEPARstart{T}{he} Chief Executive Officer (CEO) problem is defined by Berger \textit{et al.} for distributed source coding of multi-observations of a source corrupted by independent noises \cite{Berger96}. By using the compressed observations, a fusion center makes an estimation of the source at the receiver with an acceptable distortion between the original and the estimated symbols. In the last two decades, there has been an explosion of studies on the theoretical bounds of the transmission rate in the CEO problem in the case of noisy observations of a Gaussian source corrupted by independent additive Gaussian noises \cite{oha98,VB97,PTR04,oha12}.
This case is usually known as the quadratic Gaussian CEO problem. The CEO problem empirically emerges in wireless sensor networks, where a particular phenomenon is measured by some separate and independent sensors in a noisy environment.

A tight upper bound on the sum-rate distortion function of the quadratic Gaussian CEO problem and the optimal rate allocation scheme are provided in \cite{CZB04}. Alternatively, studies like \cite{CB08,BS09,BS05} present various coding schemes to achieve any point of the rate-distortion region of the quadratic Gaussian CEO problem. Moreover, an optimal coding scheme based on the successive Wyner-Ziv coding structure is applied to achieve the bounds of the quadratic Gaussian CEO in \cite{CB08}.

The case of a binary source with observations corrupted by binary noises, called the binary CEO problem, has been paid less attention during these years. In general, the exact rate-distortion bound of this case and its associated multi-terminal source coding problem are open problems in information theory. The most common criterion for measuring distortion in the binary case is the Hamming distortion measure \cite{Tad}. The binary CEO problem appears in cooperative digital communication networks where some correlated remote sources are being sent to a central receiver via paralleled channels with independent noises.

A lower bound for the rate-distortion region of a two-link binary CEO problem is established in \cite{Tad} using the Hamming distortion benchmark. The Berger-Tung inner and outer bounds \cite{ELG} are exploited for this case which are not tight under the Hamming distortion criterion. Some useful bounds on the rate-distortion performance of the binary CEO problem under the Hamming distortion measure are given in \cite{Tad2} and \cite{Tad3}. The prior studies on the binary CEO in \cite{Tad,Tad2,RA14,HBP08} consider that the correlated observations are transmitted through AWGN channels, and hence their encoders apply a channel coding to protect the transmitted data. Thus, the problem definition in those papers differs from the standard CEO problem, defined in \cite{Berger96}, for which the transmission links are assumed to be noiseless and the encoders employ source coding schemes, alternatively. In contrast, we follow the lossy distributed source coding framework in the binary CEO problem. Thus, our goal is to achieve the maximum compression of the correlated noisy observations for sending them through noiseless channels with minimum distortion.

Due to increasing demand for developing deep learning in upcoming complex networks, the logarithmic loss, or simply log-loss, has emerged as a useful criterion to measure distortion in many applications like machine learning, classification, and estimation theory. In this paper, we focus on the binary CEO problem under the log-loss criterion. This loss has been interpreted as the conditional entropy and the estimated symbols of the fusion center are soft data under this loss. Moreover, it has been also shown that the log-loss is a universal criterion for measuring the performance of lossy source coding \cite{AW15}, \cite{SRV17}. The entire achievable rate-distortion region of a two-encoder multi-terminal source coding and an $m$-encoder CEO problem under the log-loss have been derived in \cite{CW14}. The advantage of working on the CEO problem under the log-loss is that the corresponding rate-distortion region is known. By using these exact theoretical bounds, the rate-distortion performance of a designed coding scheme would be measured with more accuracy.

Our main contributions in this paper can be considered in the contexts of both information theory and coding theory. First, an exact rate-distortion bound is derived for a two-link binary CEO problem under the log-loss distortion. Next, we assume a binary symmetric channel (BSC) being used as test-channel of lossy encoders in the binary CEO problem. Then, we obtain the optimal values of crossover probabilities of the test-channels for each BSC. Finally, efficient encoding and decoding schemes are proposed by utilizing the compound LDGM-LDPC codes and iterative message-passing algorithms. We show that the rate-distortion performance of the proposed coding scheme is close to the theoretical bounds. %Moreover, the designed coding scheme is also compared to the case that Hamming distortion measure is used.

The organization of this paper is as follows. In Section II, the problem definition, preliminaries, and notations are provided. Information theoretic aspects of the binary CEO problem under the log-loss are described in Section III. Optimal values of the test-channel parameters are also presented in this section. Next in Section IV, we provide the designed encoding and decoding scheme in details. Numerical results and discussions are presented in Section V. Finally, Section VI draws the conclusion and future research.

\section{Preliminaries}

In this paper, we use uppercase letters for denoting a random variable like one used in \cite{CW14}. The realization of random variables are denoted by lowercase letters and the alphabet sets of random variables are denoted by calligraphic letters. Throughout this paper, the logarithm is to base $2$. In the Tanner graph representation of codes, first subscript shows the index of each associated link for any length, rate, distortion, and etc. Some other used notations are as follows: $p*d=p(1-d)+d(1-p)$ is binary convolution of $d$ and $p$, for $0\leq p,q \leq 1$ and $[x]^+=\max\{0,x\}$. Let $h_b(x)=-x \log x -(1-x) \log (1-x)$ be the binary entropy function where its the first and the second derivatives are, respectively, $h'_b(x)=\log \big({{1-x} \over x}\big)$ and $h''_b(x)=-{\log e \over x(1-x)}$, where $e \approx 2.7182$. The functions $h_b(x)$, $h'_b(x)$, and $h''_b(x)$ are, respectively, increasing, decreasing, and increasing functions in $x \in (0,0.5]$.

\subsection{System Model and Definitions}
\begin{figure}[t]
	\begin{center}
		\centering
		\includegraphics[width=3in,height=1.4in]{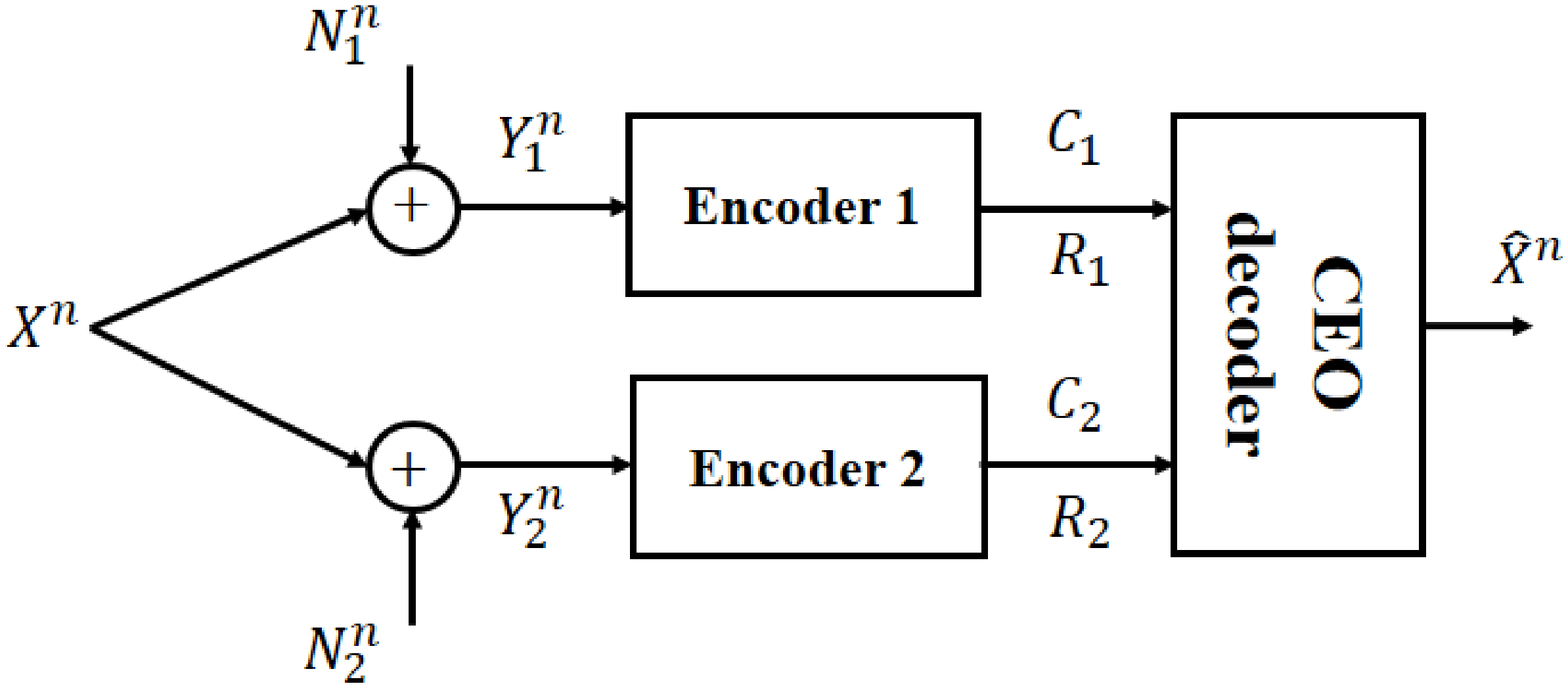}
	\end{center}
	\vspace{-20pt}
	\caption{\small{Block diagram of the two-link binary CEO problem.}}
	\label{CEO}
\end{figure}

Consider a communication system consisting of an independent and identically distributed (i.i.d.) binary symmetric source (BSS) and its two noisy observations being transmitted via two parallel links as depicted in Fig. \ref{CEO}. Let ${X}^n$, ${Y_1}^n$, and ${Y_2}^n$ denote a sequence of the BSS and two noisy observations of it, on the first and the second links, respectively. Observation noises ${N}_1^n$ and ${N}_2^n$ are independent from each other and are i.i.d. binary sequences generated by Bernoulli distributions with crossover parameters $p_1$ and $p_2$ associated to the first and the second links, respectively. Consider ${Y_1}^n$ and ${Y_2}^n$ are encoded to ${C_1}$ and ${C_2}$, and then they are sent to the CEO joint decoder. Note that ${C}_1 \leftrightarrow {Y}_1^n \leftrightarrow {X}^n \leftrightarrow {Y}_2^n \leftrightarrow {C}_2$ form a Markov chain. At the decoder, the binary sequence ${\hat X}^n$ is reconstructed in the joint decoder of CEO by using $({C}_1,{C}_2)$.

Each encoder of both links consists of a function $f_i$, $i=1,2$, which compresses the observation as follows:
\begin{equation}
\label{eqf}
f_i ({Y}_i^n)={C}_i,\, \text{where}\;  {Y}_i^n \in \mathcal{Y}_i^n = \{0,1\}^n\; \text{and} \; {C}_i \in \mathcal{C}_i ,\, \text{for}\; i=1,2.
\end{equation}
The CEO decoder is a function $g$ which maps the ordered pair $({C}_1,{C}_2)$ to the reconstruction $\hat {{X}}^n$,
\begin{equation}
\label{eqg}
g ({C}_1,{C}_2)=\hat{{X}}^n, \, \text{where} \; ({C}_1,{C}_2) \in \mathcal{C}_1 \times \mathcal{C}_2.
\end{equation}

In the lossy source coding theory, Hamming distance is a prevalent and a classic criterion for measuring the average number of flipped bits between the estimated binary sequence $\hat{{X}}^n$ compared to the original binary sequence ${X}^n$, and is denoted by $d_{\text{H}}({X}^n, \hat{{X}}^n) \delequal {1 \over n} \sum_{j=1}^{n} [x_j \oplus \hat{x}_j]$,  where $\oplus$ means the binary sum operation. If the estimated sequence may not necessarily be binary, another criterion is needed to measure the distance between these two sequences with different alphabets. In this case, an efficient conditional entropy-based distortion measure is used where probability distributions of the original source alphabet, binary in this paper, is the same as the one that will be used in the reconstructed source alphabet. This distortion measure is called log-loss.

\textit{Definition 1:} Symbol-wise log-loss between a source symbol $x_j$ and its reconstruction $\hat{x}_j$ is defined as follows:
\begin{equation}
\label{eqlog}
d(x_j , \hat{x}_j) = \log \big({1 \over \hat{x}_j (x_j)}\big), \quad j=1,2,\cdots,n.
\end{equation}
where $\hat{x}_j(x_j)$ generally depends on $({c}_1,{c}_2)$. The total value of log-loss distortion between ${x}^n$ and $\hat {{x}}^n$ is obtained by averaging over all the symbols, i.e.,
\begin{equation}
\label{eqlogg}
d({x}^n , \hat{{x}}^n) ={1 \over n} \sum_{j=1}^{n} \log \big({1 \over \hat{x}_j (x_j)}\big).
\end{equation}
%The concept of achievable rate-distortion region for the binary CEO problem is defined as follows:

\textit{Definition 2:} A rate distortion vector $(R_1,R_2,D)$ is called strict-sense achievable if there exist functions $f_1$, $f_2$, and $g$ according to (\ref{eqf}) and (\ref{eqg}) such that for length $n$,
\begin{align}
\label{eqSTA}
R_i &\ge {1 \over n} \log \big |\mathcal {C}_i \big |, \, \text{for} \; i=1,2; \\ \nonumber 
D &\ge \mathbb {E} d({X}^n, \hat{{X}}^n),
\end{align}
where $\mathbb {E}(.)$ denotes an expectation function.

\textit{Definition 3:} The closure of the set of all strict-sense achievable vectors $(R_1,R_2,D)$ is called achievable rate-distortion region of the binary CEO problem and is denoted by $\overline {\mathcal {RD}}^{\star}_{\text{CEO}}$. Furthermore, ${\mathcal {RD}}^{i}_{\text{CEO}}$ and ${\mathcal {RD}}^{o}_{\text{CEO}}$ denote the inner and the outer bounds of the rate-distortion region, respectively.

\subsection{Message-Passing Algorithms}

In our proposed coding scheme, we apply different types of message-passing algorithms depending on their applications. The Bias-Propagation (BiP) algorithm \cite{FILL07} is applied for lossy compression of a given binary source. It maps each output sequence of the source to a codeword of a given low-density generator matrix (LDGM) code which has the nearest Hamming distance to the source output. It achieves the rate-distortion bound of the BSS, and hence it is usually known as a binary quantizer in the context of source coding. Specifically, it can approach a target binary rate-distortion pair $(R,D)=(1-h_b(d),d)$ by employing LDGM codes.
In each round of this algorithm, a bias value for each variable node is calculated and then it is compared with a threshold. Regarding this comparison, the values of at least one of the variable nodes is determined in the quantized sequence. This process continues until values of all the variable nodes are fixed. Details of the BiP algorithm including update equations and damping process are presented in \cite{FILL07} and \cite{NAA17}.

Another useful message passing algorithm is the Sum-Product (SP) algorithm \cite{urban08} that is basically a decoding technique for low-density parity-check (LDPC) codes with an specified code rate and a degree distribution. In distributed lossless source coding, this algorithm is widely used as a syndrome decoder for finding the nearest sequence to a particular sequence called side information using the given nonzero syndrome, cf., \cite{CSV03} and \cite{PR03}. The iterative routine for executing this algorithm is given in \cite{LXG02}. This algorithm can asymptotically achieve the zero bit-error-rate (BER) for a target code rate equal to the capacity of a virtual channel between the original sequence and the side information.

\section{Information Theory Perspective}
In this section, we investigate the information theoretic aspect of the binary CEO problem. We review existent bounds on the rate-distortion performance of this problem and then find an optimal model for realization of them. The Berger-Tung inner and outer bounds \cite{ELG} are not generally tight, especially in the binary CEO problem case with Hamming distortion measure. If there exists a gap between the inner and the outer bounds, then measuring and comparing the rate-distortion performance of the designed codes are inaccurate. Therefore, the existence of a tight bound seems crucial for the performance analysis of a code design. Because of this, the Berger-Tung coding scheme is not optimal for the binary CEO problem under the Hamming loss in the sense of achieving the exact rate-distortion bound \cite{ELG}. In our proposed coding scheme, the total distortion is measured by using the log-loss definition (\ref{eqlogg}).

\subsection{Binary CEO Problem Bound under the Log-Loss}
Theoretical rate-distortion bound of the binary CEO problem is unknown when distortion measure is the Hamming distance, however, the inner and the outer bounds are only available for this case. Alternatively, if the log-loss criterion being used to measure distortion, the rate-distortion region is exactly established. Specifically, the classical Berger-Tung scheme yields the following inner bound of $\overline {\mathcal {RD}}^{\star}_{\text{CEO}}$. Let $(R_1,R_2,D) \in \mathcal{RD}_{\text{CEO}}^i$, if and only if, there exists a joint distribution of the form
\begin{equation}
\label{jeq}
p(x) p(y_1 | x) p(y_2 | x) p(u_1 | y_1 , q) p(u_2 | y_2 , q) p(q),
\end{equation}
where $|\mathcal{U}_i | \le |\mathcal{Y}_i  |$ for $i=1,2$, and $| \mathcal{Q} | \le 4$, which satisfies
\begin{align}
\label{innerlog}
R_1 &\ge I({Y}_1 ; {U}_1 | {U}_2 , Q), \\ \nonumber
R_2 &\ge I({Y}_2 ; {U}_2 | {U}_1 , Q), \\ \nonumber 
R_1 + R_2 &\ge I({Y}_1 , {Y}_2 ; {U}_1 , {U}_2 | Q), \\ \nonumber 
D  &\ge H({X} | {U}_1 , {U}_2 , Q).
\end{align}
Furthermore, an outer bound is provided for the binary CEO problem under the log-loss according to \textit{Definition 4} and \textit{Theorem 2} in \cite{CW14}. Let $(R_1,R_2,D) \in \mathcal{RD}_{\text{CEO}}^o$, if and only if, there exists a joint distribution of the form (\ref{jeq}) satisfies the following inequalities,
\begin{align}
\label{outerlog}
R_1 &\ge [I({Y}_1 ; {U}_1 | {X} , Q)+H({X}|{U}_2,Q)-D]^+, \\ \nonumber
R_2 &\ge [I({Y}_2 ; {U}_2 | {X} , Q)+H({X}|{U}_1,Q)-D]^+, \\ \nonumber 
R_1 + R_2 &\ge [I({Y}_1 ; {U}_1 | {X} , Q)+I({Y}_2 ; {U}_2 | {X} , Q)+H({X})-D]^+, \\ \nonumber
D  &\ge H({X} | {U}_1 , {U}_2 , Q).
\end{align}
The most important result in \cite{CW14}, related to our work, is \textit{Theorem 3} and its extension. It states that the bounds in (\ref{innerlog}) and (\ref{outerlog}) are the same, yielding a complete characterization of the rate-distortion region under the log-loss. Therefore, the Berger-Tung inner bound is tight under the log-loss. We focus on the two-link binary CEO problem, however it can be extended to $m$-link case. 
By applying the well-known \textit{support lemma} \cite{CW14} for attaining the cardinality bound, all rate-distortion vectors can be achieved in $\overline{\mathcal{RD}}_{CEO}^{\star}$ if $| \mathcal{Q} | \le 4$ and $|\mathcal{U}_i | \le |\mathcal{Y}_i  | = 2$, for $i=1,2$, are satisfied.
%Therefore, the most general case for the test-channel model of the encoders between $\bm{Y}_i$ and $\bm{U}_i$ is a $2 \rightarrow 4$ channel with transition matrix $\bm{T}^{(i)}=[t_{k,j}];\ i=1,2$ and $j=1,2,3,4$. Additionally, consider $\bm{T}^{(i)}_F=[t_{k,5-j}]$ to be its flipped transition matrix associated to the flipped test-channel. 
To find a complete characterization of the sum-rate-distortion function for the two-link binary CEO problem, we should solve the following optimization problem:
\begin{align}
\label{opt1}
\underset{p(u_1|y_1,q)p(u_2|y_2,q)p(q)}{\text{min}}& \ \ I({U}_1,{U}_2;{Y}_1,{Y}_2 | Q), \\ \nonumber
\text{s.t.}& \ \  H({X}|{U}_1,{U}_2 , Q)= D_0,
\end{align}
where $H({X}|{Y}_1,{Y}_2) \le D_0 \le 1$. %The rate and distortion functions remain unchanged under the ``flipping'' procedure, i.e., $R_F=R$ and $D_F=D$. Next, let consider the mixed channel $\lambda \bm{T}^{(i)} + (1-\lambda) \bm{T}^{(i)}_F$ for each link $i=1,2$. Obviously, the mixed channel for $\lambda=0.5$ is a symmetric channel.
%\textit{lemma :} It is shown by utilizing the support lemma that the general case of test-channels for a $l$-link binary CEO problem is a $2 \rightarrow 2+2^{l-1}$ channel.
%\textit{Lemma 1:} A $2 \rightarrow 4$ symmetric channel is equivalent with a $2 \rightarrow 2$ symmetric channel.
%\begin{proof}[Proof]
%Proof is done by using the above mentioned idea. Output $U$ with alphabet $\{0,1,2,3\}$ can be considered as an ordered pair $U=(V,W)$, where $V$ and $W$ are binary and they are independent from each other. Obviously, this channel is equivalent to a state space diagram that its input, output and state are, respectively, $Y$, $V$ and $W$. Because $V$ and $W$ are independent, the state has not any effect on the output, and hence the transition probabilities are added together as to lead to a BSC.
%\end{proof}
This optimization problem can be written in the following unconstrained form:
\begin{align}
\label{opt2}
\underset{p(u_1|y_1,q)p(u_2|y_2,q)p(q)}{\text{min}} \ \ H({X}|{U}_1,{U}_2 , Q) + {\mu} I({U}_1,{U}_2;{Y}_1,{Y}_2 | Q),
\end{align}
where $\mu$ is the Lagrangian multiplier. Note that 
\begin{align}
	\label{opt21}
	&H({X}|{U}_1,{U}_2 , Q) + {\mu} I({U}_1,{U}_2;{Y}_1,{Y}_2 | Q) \\ \nonumber
	&=\sum_{q \in \mathcal{Q}} p(q) [H({X}|{U}_1,{U}_2 , Q=q) + {\mu} I({U}_1,{U}_2;{Y}_1,{Y}_2 | Q=q)] \\ \nonumber
	& \ge \underset{q \in \mathcal{Q}}{\text{min}} \ \ H({X}|{U}_1,{U}_2 , Q=q) + {\mu} I({U}_1,{U}_2;{Y}_1,{Y}_2 | Q=q).
\end{align}
Therefore, for the purpose of characterizing the sum-rate-distortion function, there is no loss of generality in assuming that $Q$ is a constant, which leads to the following simplified optimization problem:
\begin{align}
\label{opt22}
\underset{p(u_1|y_1)p(u_2|y_2)}{\text{min}} \ \ H({X}|{U}_1,{U}_2) + {\mu} I({U}_1,{U}_2;{Y}_1,{Y}_2) \delequal F.
\end{align}
%\textit{Lemma :} For the optimization problem (\ref{opt2}), and for any asymmetric channel $C$, there is another test-channel model with smaller value of the objective function $F$ for any $\mu \neq \mu_0$, where $\mu_0$ is a special value depending on the noise parameters $p_1$ and $p_2$.
%\textit{Proof:} Consider linear combination $\lambda C+ (1-\lambda) C_F$ for any $0 \le \lambda \le 1$, this channel is a $2 \rightarrow 4$ channel that is $C$ and $C_F$ when $\lambda=1$ and $\lambda=0$, respectively. Furthermore, this channel is equivalent with a symmetric channel in $\lambda=0.5$. Therefore, we shall to study behavior of the objective function with respect to the parameter $\lambda$. ..... 
\subsection{BSC Test-Channel Model for the Encoders}
We shall assume that $p(u_i | y_i )$ is a BSC with crossover probability $d_i, \ i =1,2$, which is justified by the extensive numerical solutions to (\ref{opt22}).
%Based on numerical results for solving (\ref{opt22}) and finding the optimal test-channel model for the encoders of the first and the second links, they are assumed to be BSC with crossover probabilities $d_1$ and $d_2$, respectively. In other words, we further restrict $p(u_i | y_i)$ to be a BSC with crossover probability $d_i$, for $i=1,2$.
%By this assumption, the block diagram of the binary CEO problem model is shown in the Fig. \ref{CEO-BSC}.
%\begin{figure}[t]
%	\begin{center}
%		\centering
%		\includegraphics[width=3in,height=1.2in]{CEO-BSC.eps}
%	\end{center}
%	\vspace{-20pt}
%	\caption{The binary CEO problem with BSC test-channel models.}
%	\label{CEO-BSC}
%\end{figure}
Consequently, after some calculus manipulations in (\ref{innerlog}), the rate-distortion bounds are expressed by:
\begin{align}
\label{eq3}
R_1 &\ge h_b(p*d)-h_b(d_1), \\ \nonumber
R_2 &\ge h_b(p*d)-h_b(d_2), \\ \nonumber
R \delequal R_1 + R_2 &\ge 1+h_b(p*d)-h_b(d_1)-h_b(d_2),  \\ \nonumber
D  &\ge h_b(p_1*d_1)+h_b(p_2*d_2)-h_b(p*d),
\end{align}
where $p \delequal p_1*p_2$ and $d \delequal d_1*d_2$. In this case, the optimization problem (\ref{opt22}) is equivalent to:
\begin{align}
\label{opt3}
\underset{0 \le d_1,d_2 \le 0.5}{\text{min}} \ \ h_b(p_1*d_1)+h_b(p_2*d_2)-h_b(p*d)+\mu \big(1+ h_b(p*d)-h_b(d_1)-h_b(d_2) \big),
\end{align}
for any $\mu$. The following example gives an intuition that $d_1=d_2$ in the binary CEO problem is not necessarily optimum choice even if $p_1=p_2$.

\textit{Example 1:} Let consider $p_1=p_2=0.1$ and also assume that the minimum achievable sum-rate $R$ is fixed to $0.6$, i.e., $1+h_b(p*d)-h_b(d_1)-h_b(d_2)=0.6$. First, let $d_1=d_2$. By a simple calculation, $d_1=d_2=0.177$ is obtained, and then the minimum achievable distortion $D$ will be equal to $0.6474$.
Alternatively, let presume that the total information is only sent over the first link, i.e., $d_2=0.5$. Thus, $d_1$ and $D$ are calculated as $0.0795$ and $0.6428$, respectively. Consequently, the distortion value by using only one of the links is unexpectedly smaller than the case that both of them are used, and hence finding optimum values of $d_1$ and $d_2$ is interesting. Optimality in this case means achieving the minimum achievable log-loss distortion subject to a given minimum achievable sum-rate. First of all, we show that the optimization problem (\ref{opt1}) is not convex even with the BSC assumption for the encoders.

\textit{Theorem 1:} Bounds of distortion $D$ and sum-rate $R$ in (\ref{eq3}) are neither convex nor concave in terms of variables $(d_1,d_2)$. 

\begin{proof}[Proof]
	From (\ref{eq3}) we have:
	\begin{align}
	\label{eq4}
	\frac{\partial^2 R}{\partial d_i^2} &=(1-2p*d_{3-i})^2h''_b(p*d)-h''_b(d_i), \\ \nonumber 
	\frac{\partial^2 R}{\partial d_1 \partial d_2} &= (1-2p*d_1)(1-2p*d_2)h''_b(p*d)-2(1-2p)h'_b(p*d),\\ \nonumber
	\frac{\partial^2 D}{\partial d_i^2}&=(1-2p_i)^2h''_b(p_i*d_i)-(1-2p*d_{3-i})^2h''_b(p*d), \\ \nonumber \frac{\partial^2 D}{\partial d_1 \partial d_2} &= -(1-2p*d_1)(1-2p*d_2)h''_b(p*d)+2(1-2p)h'_b(p*d).
	\end{align}
	The Hessian matrices of the rate and distortion are, respectively, ${H}_R=[\frac{\partial^2 R}{\partial d_i\partial d_j}]$ and ${H}_D=[\frac{\partial^2 D}{\partial d_i\partial d_j}]$, for $i,j=1,2$. By defining $q_i \delequal p*d_{3-i}$, obviously $q_i \ge p_i$. After some calculations, we have:
	\begin{align}
	\label{Hesscal}
	&\frac{\partial^2 R}{\partial d_i^2} ={q_i(1-q_i) \over q_i*d_i(1-q_i*d_i)d_i(1-d_i)} \ge 0, \quad \frac{\partial^2 D}{\partial d_i^2}={p_i(1-p_i)-q_i(1-q_i) \over p_i*d_i(1-p_i*d_i)q_i*d_i(1-q_i*d_i)} \le 0.
	\end{align}
	For fixed values of $d_i$, sum-rate $R$ and distortion $D$ are, respectively, a convex and a concave single-variable functions in terms of $d_{3-i}$, for $i=1,2$ according to (\ref{Hesscal}).
	We show that the determinant of the Hessian matrices for $R$ and $D$ are not positive with a counter-example.
	%\textit{Lemma 3:} For any $x \ge 0$, $\log(1+x) \le x $, and therefore for any $0 \le y \le 0.5$, let $1+x\delequal {1-y \over y} \ge 1$, then $\log\big({1-y \over y}\big) \le {1-2y \over y}$. 
	%\textit{Lemma 4:} If $x_1 \ge y_1 \ge 0$ and $x_2 \ge y_2 \ge 0$, then $x_1x_2+y_1y_2 \ge x_1y_2+x_2y_1$.
	%\textit{Proof:} We have, 
	%\begin{align}
	%(x_1x_2+y_1y_2)-(x_1y_2+x_2y_1)= x_1(x_2-y_2)-y_1(y_2-x_2)=(x_1-y_1)(x_2-y_2) \ge 0. \qed
	%\end{align}
	%\textit{Lemma 5:} For any $0 \le p \le 0.5$ and $0 \le d_1,d_2 \le 0.5$ let $d=d_1*d_2$, then:
	%\begin{equation}
	%\label{ineq}
	%\Scale[0.9]{(1-2p)(1-p*d)(1-2p*d) \ge {\sqrt 2-1 \over 2}(1-2p*d_1)(1-2p*d_2).}
	%\end{equation}
	%\textit{Proof:} Let define $f(d_1)=(1-2p)(1-p*d)(1-2p*d) - {\sqrt 2-1 \over 2}(1-2p*d_1)(1-2p*d_2)$, clearly $f(0.5)=0$. We prove $f'(d_1)<0$ for $0 \le d_1 < 0.5$, therefore $f(d_1) \ge 0$.
	%\begin{align}
	%&\Scale[0.8]{f'(d_1)=(1-2p)\big[-(1-2p*d_2)(1-2p*d)-2(1-p*d)(1-2p*d_2) \big] + (\sqrt 2-1)(1-2p)(1-2p*d_2)} \\ \nonumber
	%&\Scale[0.8]{\equiv -(1-2p*d)-2(1-p*d) + (\sqrt 2-1)=-4+\sqrt 2+4p*d=-2.5858+4p*d \le -2.5858+2=-0.5858,}
	%\end{align}
	%where $\equiv$ shows equivalency in sign. The determinant of Hessian matrices after some simplifications and using lemma $3$ in the expression $h'_b(p*d)$ are as follows,
	\begin{align}
	\label{eqhee}
	&\det\big[{H}_R\big]=\frac{\partial^2 R}{\partial d_1^2}\frac{\partial^2 R}{\partial d_2^2}-\big(\frac{\partial^2 R}{\partial d_1\partial d_2}\big)^2 \\ \nonumber &=\big( {q_1 (1-q_1) \over p*d(1-p*d)d_1(1-d_1)}\big) \times  \big({ q_2(1-q_2) \over p*d(1-p*d)d_2(1-d_2)}\big) \\ \nonumber &- \big(2(1-2p) \log \big[{1-p*d \over p*d}\big]+{(1-2q_1)(1-2q_2) \over p*d(1-p*d)}\big)^2.
	\end{align}
	Let calculate $\det\big[{H}_R\big]$ for $d_1=d_2=0.1$ where $p_1 \to 0$ and $p_2 \to 0$. In this case, $d=0.18$, $p \to 0$, $q_1 \to 0.1$, and $q_2 \to 0.1$. Hence, 
	\begin{align}
	\label{ceqhee}
	&\det\big[{H}_R\big]=\big( {0.09 \over 0.0133}\big) \times  \big({ 0.09 \over 0.0133}\big)- \big(3.0327+4.3360\big)^2=-8.5066<0.
	\end{align}
	This counter-example shows that $R$ is neither convex nor concave in general. Similarly, for distortion $D$, we have:
	\begin{align}
	\label{eqhe}
	&\det\big[{H}_D\big]=\frac{\partial^2 D}{\partial d_1^2}\frac{\partial^2 D}{\partial d_2^2}-\big(\frac{\partial^2 D}{\partial d_1\partial d_2}\big)^2 \\ \nonumber &=\big( {q_1 (1-q_1)-p_1(1-p_1) \over p*d(1-p*d)p_1*d_1(1-p_1*d_1)}\big) \times  \big({ q_2(1-q_2)-p_2(1-p_2) \over p*d(1-p*d)p_2*d_2(1-p_2*d_2)}\big) \\ \nonumber &- \big(2(1-2p) \log \big[{1-p*d \over p*d}\big]+{(1-2q_1)(1-2q_2) \over p*d(1-p*d)}\big)^2.
	\end{align}
	Now we calculate $\det\big[{H}_D\big]$ for $p_1=p_2=0.1$ when $d_1 \to 0$ and $d_2 \to 0$. In this case, $p=0.18$, $d \to 0$, $q_1 \to 0.18$, and $q_2 \to 0.18$. Hence, 
	\begin{align}
	\label{ceqhe}
	&\det\big[{H}_D\big]=\big( {0.0576 \over 0.0133}\big) \times  \big({ 0.0576 \over 0.0133}\big)- \big(1.9409+2.7751\big)^2=-3.4846<0.
	\end{align}
	This counter-example also shows that $D$ is neither convex nor concave in general.
\end{proof}

%where,
%\begin{align}
%\label{AB}
%&\Scale[0.9]{A \delequal 4(1-2p)^2(1-p*d)^2(1-2p*d)^2(p_1*d_1)(1-p_1*d_1)(p_2*d_2)(1-p_2*d_2),} \\ \nonumber
%&\Scale[0.9]{B \delequal 4(1-2p)(1-p*d)(1-2p*d)(1-2q_1)(1-2q_2)(p_1*d_1)(1-p_1*d_1)(p_2*d_2)(1-p_2*d_2).}
%\end{align}
%It is concluded from (\ref{ineq}) that,
%\begin{align}
%\label{ineqq}
%&\Scale[0.9]{A+B= 4(1-2p)^2(1-p*d)^2(1-2p*d)^2(p_1*d_1)(1-p_1*d_1)(p_2*d_2)(1-p_2*d_2)+}\\ \nonumber 
%&\Scale[0.9]{4(1-2p)(1-p*d)(1-2p*d)(1-2p*d_1)(1-2p*d_2)(p_1*d_1)(1-p_1*d_1)(p_2*d_2)(1-p_2*d_2)} \\ \nonumber
%&\Scale[0.9]{ \ge  (1-2p*d_1)^2(1-2p*d_2)^2(p_1*d_1)(1-p_1*d_1)(p_2*d_2)(1-p_2*d_2).}
%\end{align}
%Finally, we put (\ref{ineqq}) in (\ref{eqhe}) and take parameters in the inequality lemma $4$ such as follows:
%\begin{align}
%\label{xy}
%&\Scale[0.9]{x_1=(1-2p_1)^2(p*d)(1-p*d),} \
%\Scale[0.9]{x_2=(1-2p_1)^2(p*d)(1-p*d),} \\ \nonumber
%&\Scale[0.9]{y_1=(1-2q_1)^2(p_1*d_1)(1-p_1*d_1),} \
%\Scale[0.9]{y_2=(1-2q_2)^2(p_2*d_2)(1-p_1*d_2).}
%\end{align}
%Therefore, according to the above lemma, we proved that $\det\big[\bm{H}_D\big] \ge 0$ and distortion $D$ is concave. Furthermore, from (\ref{eq4}) it is obvious that $\frac{\partial^2 R}{\partial d_i^2}  \ge -  \frac{\partial^2 D}{\partial d_i^2} \ge 0$ and $\frac{\partial^2 R}{\partial d_id_j}  = -  \frac{\partial^2 D}{\partial d_id_j}$. Thus, $\det\big[\bm{H}_R\big] \ge \det\big[\bm{H}_D\big] \ge 0$, and thus the rate $R$ is convex and proof is completed. $\square$

Regarding the above theorem, objective function $F(\mu) \delequal D+\mu R$ in (\ref{opt3}), which is a two-dimensional function of $(d_1,d_2)$, is not convex in general. For a fixed value of $\mu$, solution of (\ref{opt3}) is an ordered pair denoted by $(d_1^*,d_2^*)$ which achieves minimum value of $F$. If $p_1=p_2$, only pairs that $d_1^* \le d_2^*$ will be considered as an acceptable solution due to the symmetry. Now, solution of the problem (\ref{opt3}) is presented by utilizing an exhaustive search on the plane $(d_1,d_2)$ with sufficiently small step-sizes.

\begin{figure}[t]
	\centering
	\subfigure[{For $p_2=p_1$.}]{\label{fig1}
		\includegraphics[width=2in,height=1.7in]{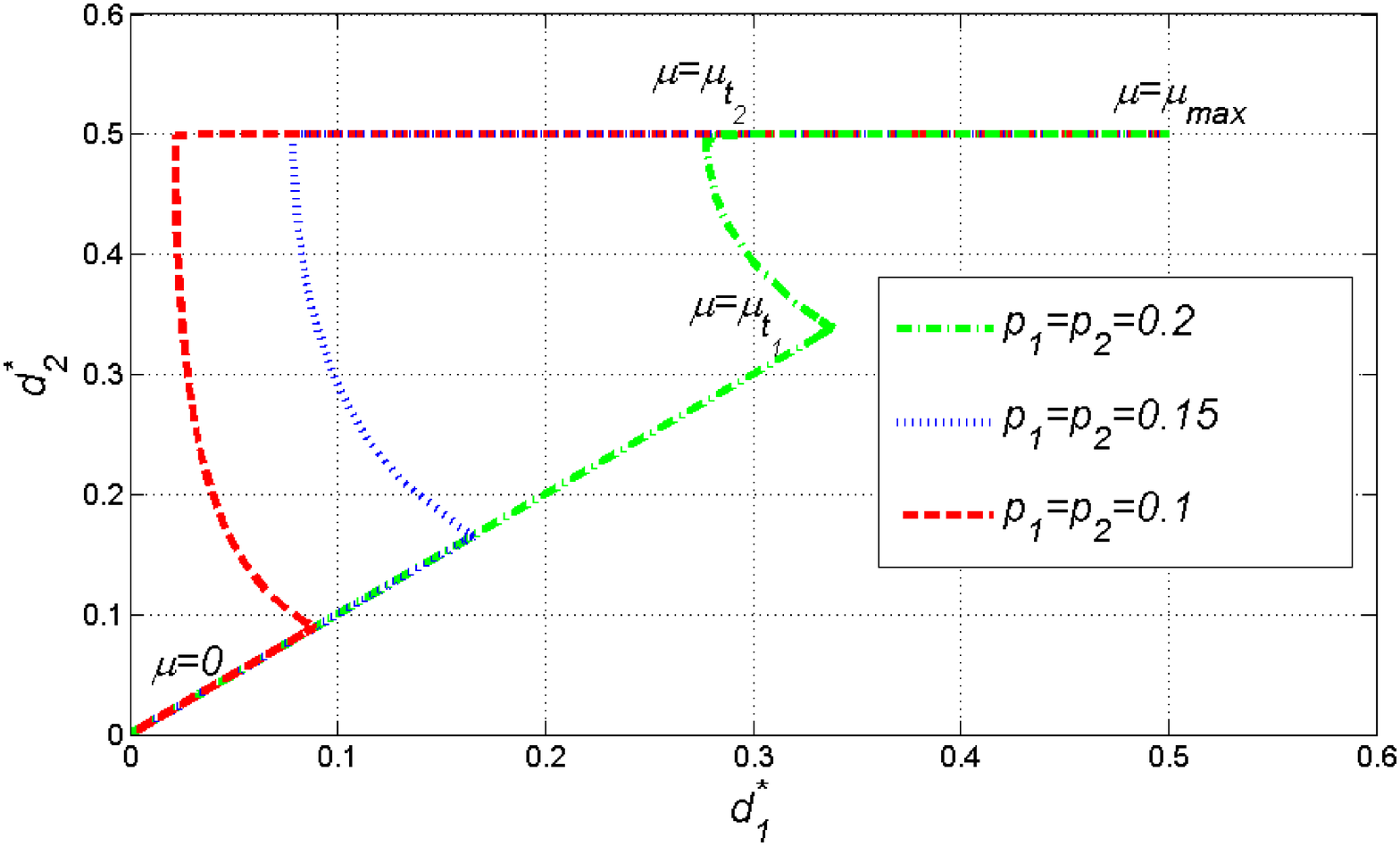}}
	\hspace{0.5mm}
	\subfigure[{For $p_2-p_1=0.01$.}]{\label{fig2}
		\includegraphics[width=2in,height=1.7in]{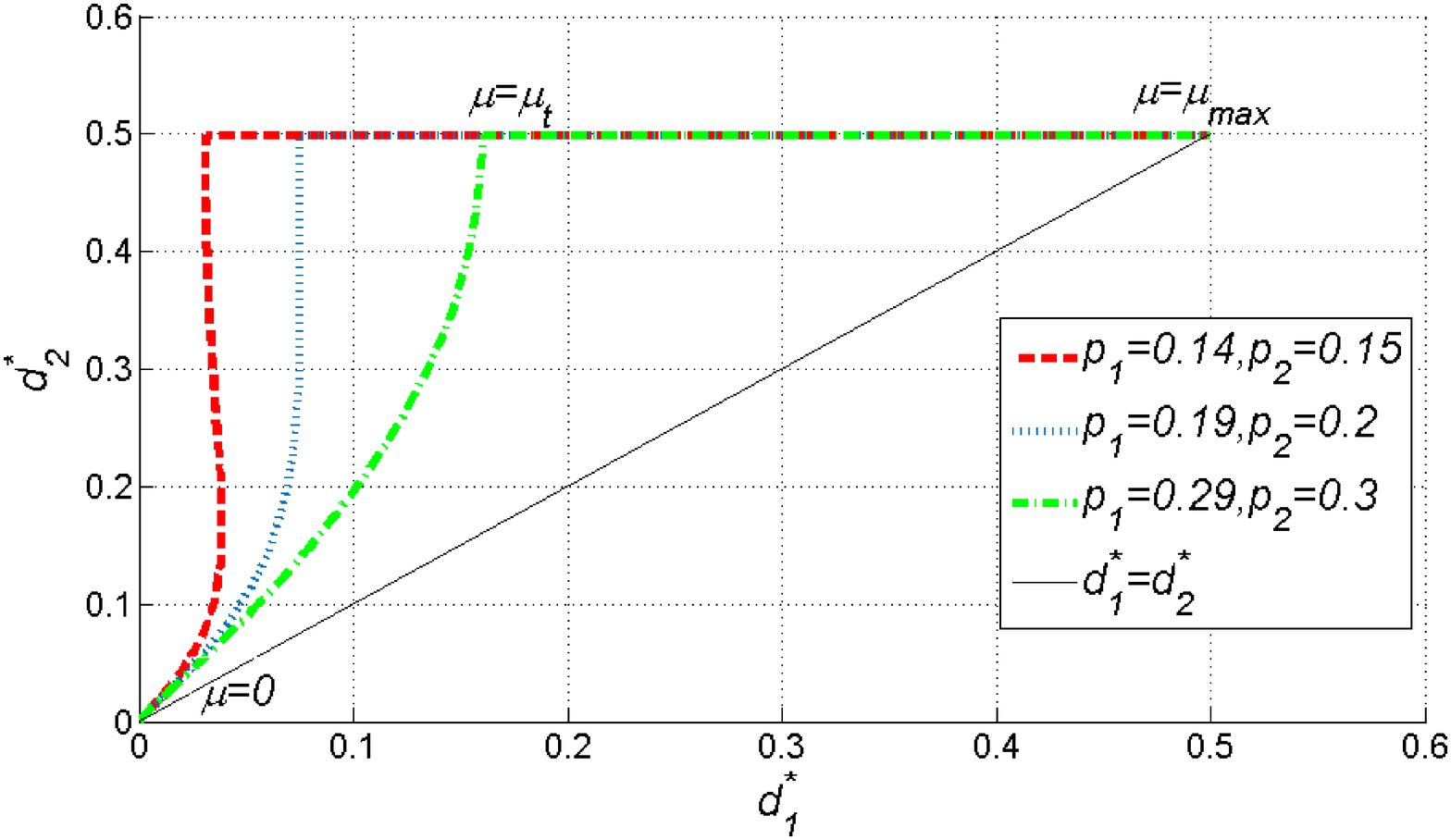}}
	\hspace{0.5mm}
	\subfigure[{For $p_2-p_1=0.05$.}]{\label{fig3}
		\includegraphics[width=2in,height=1.7in]{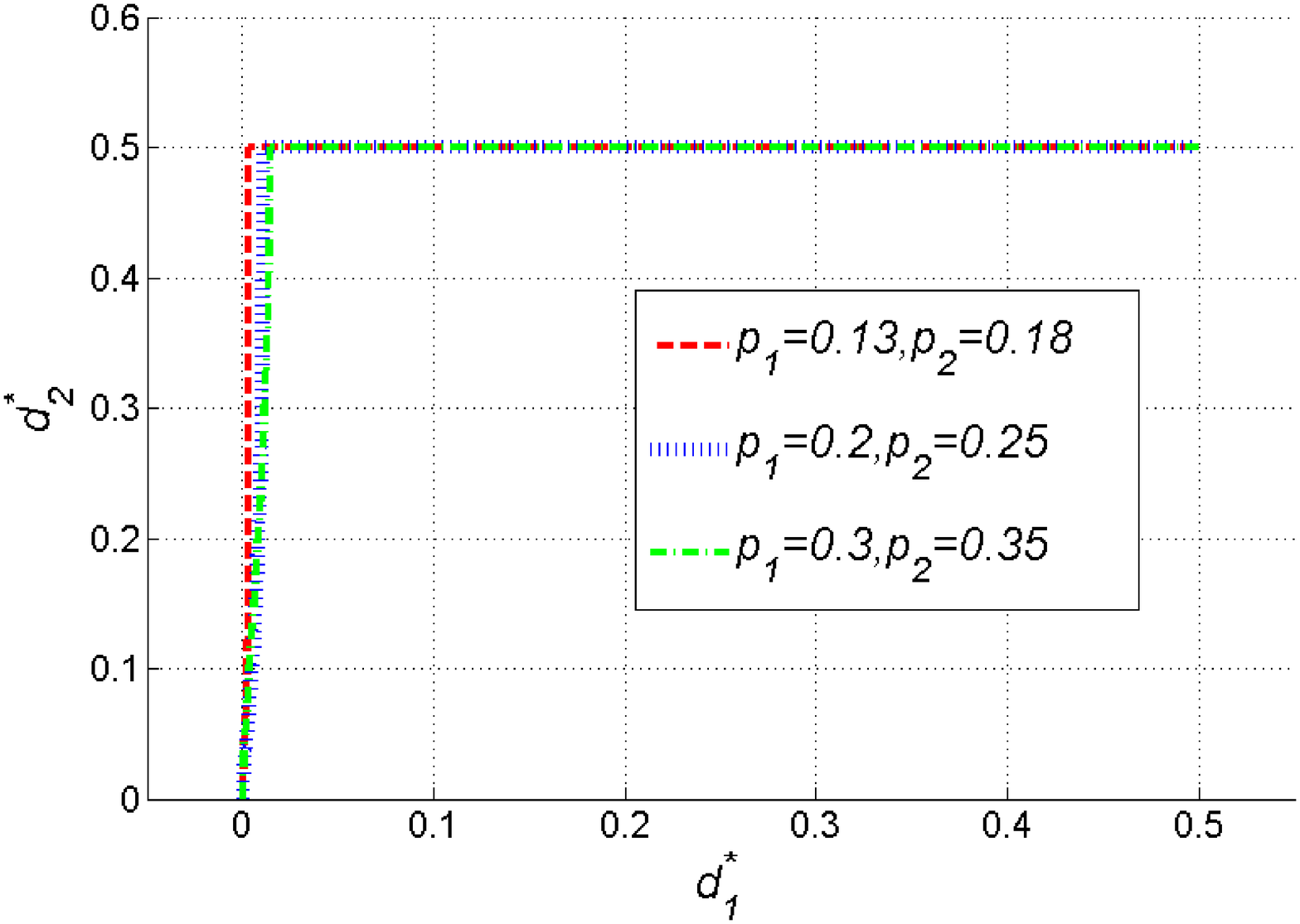}}
	\caption{Location of the optimum points $(d_1^*,d_2^*)$.}
	\label{fig123}
\end{figure}

There exist two definite boundary points in $F(\mu)$. First, it arises in $\mu=0$ where the objective function equals distortion, and hence the minimum value $D_{\text{min}}$ equals $H({X}|{Y}_1,{Y}_2)$ for $(d_1^*,d_2^*)=(0,0)$. Second, it occurs in $\mu_{\text{max}}$ by which the minimum value of the objective function equals to $H({X})=1$ for all $\mu$ values equal or greater than $\mu_{\text{max}}$, i.e., $F_{\text{min}}(\mu)=1$ for $\forall \mu \ge \mu_{max}$. In the latter boundary point, the solution is located in $(d_1^*,d_2^*)=(0.5,0.5)$ and the sum-rate $R=0$. Thus, it is sufficient to study behavior of the objective function between these two boundary points, i.e., $0 \le \mu \le \mu_{max}$.
Our results show that location of the solutions depends on the value of noise parameters $p_1$ and $p_2$ whether they are equal or not.
%\subsubsection{Equal Noise-Parameter Case, $p_1=p_2$}
In Fig. \ref{fig123}, location of the solution points $(d_1^*,d_2^*)$ are depicted for several cases. When $p_1=p_2$, as it is seen in these curves, there exist two threshold values for parameter $\mu$, denoted by $\mu_{t_1}$ and $\mu_{t_2}$, related to non-smooth critical points of the curves. These critical points are used to categorize location of the optimum solutions. In \textit{Region 1}, $0 \le \mu \le \mu_{t_1}$, the optimum points $(d_1^*,d_2^*)$ are located on the line $d_1^*=d_2^*$. In \textit{Region 2}, $\mu_{t_1} \le \mu \le \mu_{t_2}$, the optimum points are located on a curve such that $d_1^*< d_2^*<0.5$. In \textit{Region 3}, $\mu_{t_2} \le \mu \le \mu_{max}$, the solutions are located on the boundary points. However, when $p_1 \neq p_2$, there is only one threshold value for $\mu$ denoted by $\mu_t$ corresponding to the critical point of the curve. In \textit{Region 1}, $0 \le \mu \le \mu_{t}$, the solutions are located on a curve such that $d_1^* <  d_2^*< 0.5$, and in \textit{Region 2}, $\mu_t \le \mu \le \mu_{max}$, the optimum points are located on the boundary points.
As it is seen in Fig. \ref{fig123}, one of the links becomes useless for sending encoded observations to the decoder when the difference $p_2-p_1$ slightly increases. To confirm these solutions, all roots of the gradient equation ${\nabla}F=[\frac{\partial F}{\partial d_1} ,\frac{\partial F}{\partial d_2} ]=[0 \ 0]$ are calculated. These roots give all possible optimum points except the boundary points, hence:
\begin{equation}
\label{eqgradd}
(1-2p_i)h'_b(d_i*p_i)- \mu h'_b(d_i)+(\mu-1)(1-2q_i)h'_b(d*p)=0, \quad \text{for} \quad i=1,2.
\end{equation}
This non-linear system of equations dose not have any closed-form solution and it is generally solved by using numerical methods such as Newton's method \cite{Remani13}. Next, the Hessian matrix ${H}_F={H}_D+\mu {H}_R$ is calculated in these roots to check whether a possible point is exactly an optimum point or not.
\begin{equation}
\label{eq9}
{H}_F=\big[\frac{\partial^2 F}{\partial d_i \partial d_j}\big]; \ i,j \in \{1,2\}.
\end{equation}
If ${H}_R$ is a positive definite matrix in a root of ${\nabla}F=[0 \ 0]$, then it is a solution. Otherwise, since there is not such a point, the solution is only located on the boundary points, i.e., at least one of the $d_i^*$s equals $0$ or $0.5$. Due to the non-convexity of the optimization problem and the non-linearity of the rate and distortion expressions, we give an asymptotic analysis of the problem (\ref{opt3}). In this regard, we consider a high resolution regime.

\textit{Lemma 1:} Assume $K=(1-2p)\log\big({1-p \over p}\big)$ and $K_i=(1-2p_i)\log\big({1-p_i \over p_i}\big)$ for $i=1,2$. For $x \to 0$,
\begin{align}
\label{eq10}
\ h_b(x)=-x\log(x)+x+O(x^2), \ \ h_b(x*p)-h_b(p)=Kx+O(x^2).
\end{align}

\textit{Theorem 2:} Location of the solution points of (\ref{opt3}), when $d_1 \rightarrow 0$ and $d_2 \rightarrow 0$, is as follows:
\begin{equation}
\label{eq11}
d_2 \approx e^{K(K_2-K) \over (K_1-K)}d_1^{K_2-K \over K_1-K}.
\end{equation}

\begin{proof}[Proof]
	Let $R_0=1+h_b(p)$ and $D_0=h_b(p_1)+h_b(p_2)-h_b(p)$ be sum-rate and distortion at $(d_1,d_2)=(0,0)$. According to (\ref{eq10}), after $(d_1,d_2) \rightarrow (0,0)$, then we have:
	\begin{align}
	\label{eq12}
	R-R_0&=h_b(p*d)-h_b(p)-h_b(d_1)-h_b(d_2), \\ \nonumber 
	&=Kd+d_1\log d_1 -d_1 +d_2\log d_2 -d_2+O(\max \{d^2,d_1^2,d_2^2\}) \\ \nonumber
	&=(K-1)(d_1+d_2)+d_1 \log d_1 + d_2 \log d_2 +O(\max \{d_1^2,d_2^2\}),
	\end{align}
	\begin{align}
	\label{eq13}
	D-D_0&=h_b(p_1*d_1)-h_b(p_1)+h_b(p_2*d_2)-h_b(p_2)-h_b(p*d)+h_b(p) \\ \nonumber
	&=K_1d_1+K_2d_2-Kd+O(\max \{d^2,d_1^2,d_2^2\}) \\ \nonumber
	&=(K_1-K)d_1+(K_2-K)d_2+O(\max \{d_1^2,d_2^2\}).
	\end{align}
	By ignoring the high-order terms, the following convex optimization problem is obtained:
	\begin{equation}
	\begin{aligned}
	\label{opt4}
	& \underset{0 \le d_1,d_2 \le 0.5}{\text{min}}
	& &  (K-1)(d_1+d_2)+d_1 \log d_1 + d_2 \log d_2 \\
	& \text{s.t.} & &  (K_1-K)d_1+(K_2-K)d_2=D-D_0. \\
	\end{aligned}
	\end{equation}
	The Lagrangian $L$ of the above minimization problem for the Lagrangian multiplier $\lambda$ is given by:
	\begin{equation}
	\label{eq14}
	L=(K-1)(d_1+d_2)+d_1 \log d_1 + d_2 \log d_2 + \lambda [(K_1-K)d_1+(K_2-K)d_2-D+D_0].
	\end{equation}
	The gradient equation implies:
	\begin{equation}
	\label{eq15}
	{\partial L \over \partial d_i}=K+\log d_i + \lambda (K_i - K)=0, \quad \text{for} \quad i=1,2.
	\end{equation}
	Finally, by canceling $\lambda$ in the above two equations, it is concluded that:
	\begin{equation}
	\label{eq16}
	K(K_2-K_1)+(K_2-K)\log d_1 - (K_1-K)\log d_2=0 \quad \Rightarrow \quad d_2=e^{K(K_2-K) \over (K_1-K)}d_1^{K_2-K \over K_1-K}.
	\end{equation}	
\end{proof}
\textit{Corollary 1: } In general, without assuming that the test-channels are BSCs, one can still show that $R-R_0$ can be approximated by a convex function while $D-D_0$ can be approximated by a linear function, in the high-resolution regime. As a consequence, computation of the rate-distortion function can be approximately formulated as a convex optimization problem. A direct implication of this convex optimization formulation is that the binary symmetric test-channel is asymptotically optimal in the high-resolution regime. 

\textit{Corollary 2:} The slope of the tangent lines to the curve of location of optimum points in $(d_1^*,d_2^*)=(0,0)$ are, respectively, $1$, $\infty$, or $0$ when $p_1=p_2$, $p_1<p_2$, or $p_1>p_2$.

In the following figure, location of the optimum points in a high-resolution regime \footnote{$d_1$ and $d_2$ are $O(10^{-4})$} and the curve (\ref{eq11}) for Fig. \ref{fig2} are depicted. As it is obvious, these curves are approximately the same.
By the following lemma, the asymptotic analysis of the problem is investigated around $(d_1^*,d_2^*)=(0.5,0.5)$.
\begin{figure}[t]
	\centering
	\subfigure[{Depiction of Fig. \ref{fig2} in a high-resolution regime.}]{\label{fig4}
		\includegraphics[width=2.5in,height=2in]{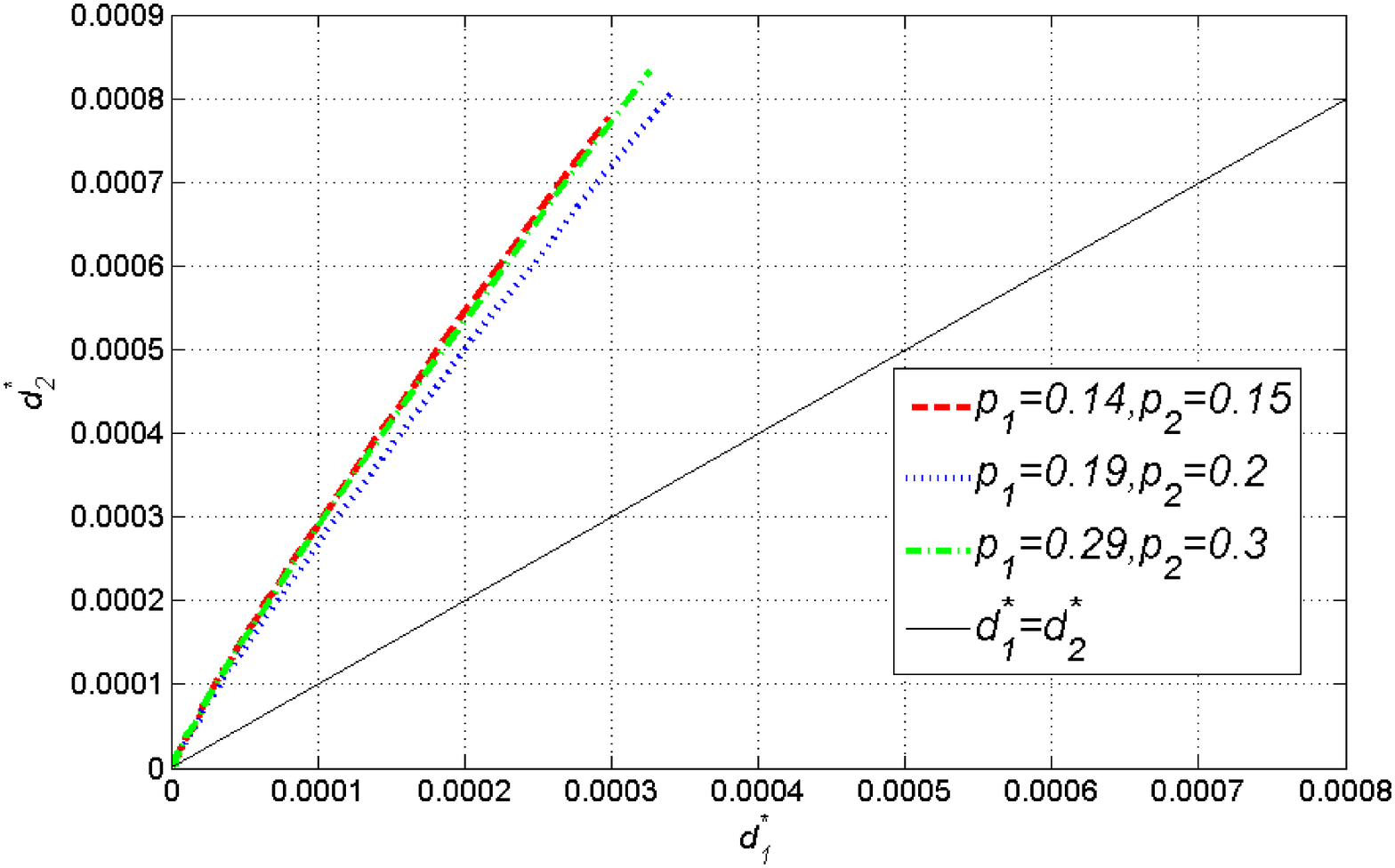}}
	\hspace{1mm}
	\subfigure[{Direct plotting of (\ref{eq11}).}]{\label{fig5}
		\includegraphics[width=2.5in,height=2in]{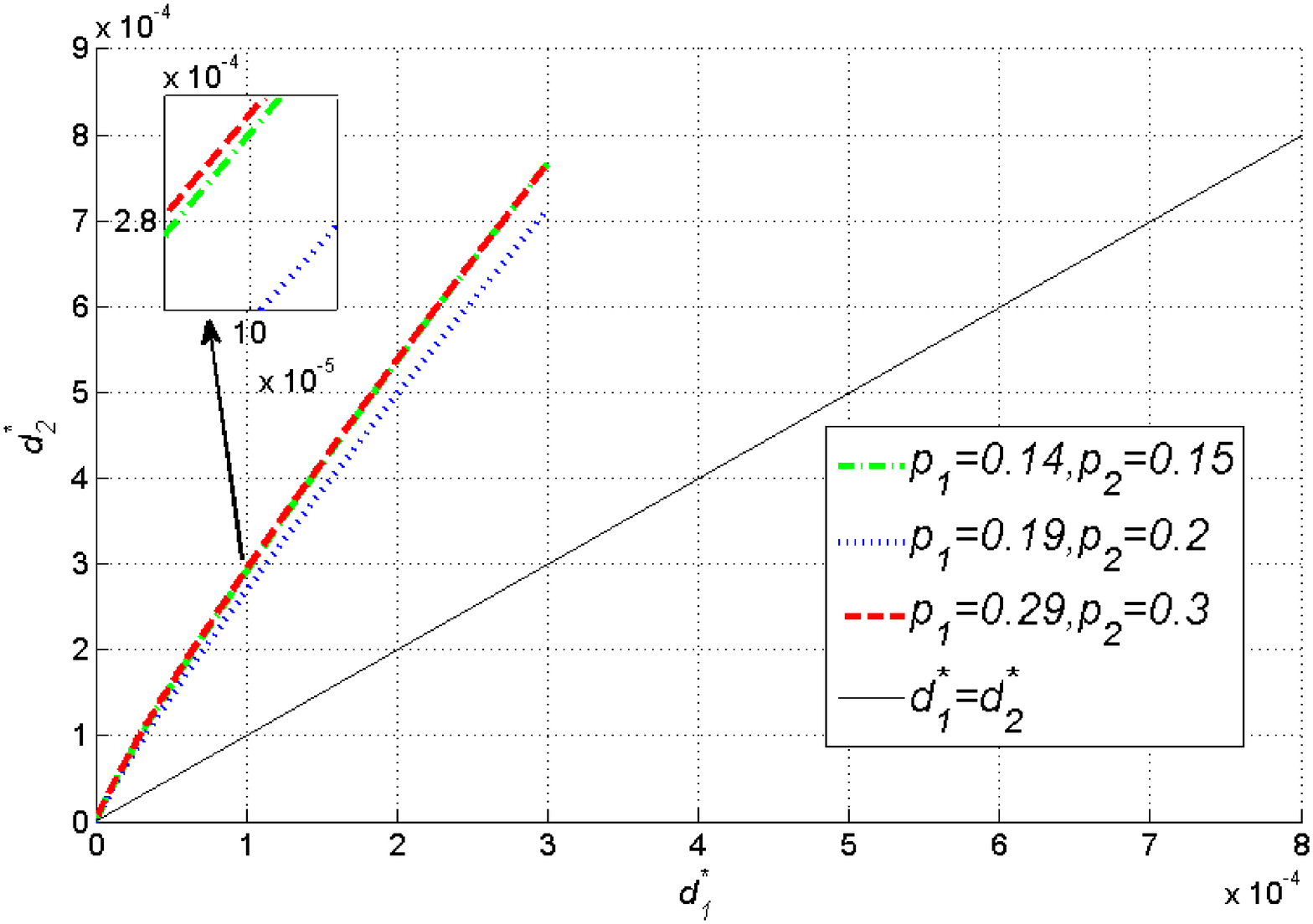}}
	\caption{Comparison of the location of optimum points and curves of (\ref{eq11}).}
	\label{fig45}
\end{figure}

\textit{Lemma 2:} The maximum value of the parameter $\mu$ occurs in $(R,D)=(0,1)$ when $(d_1^*,d_2^*)=(0.5,0.5)$ and it equals:
\begin{equation}
\label{mu}
\mu_{\max}= \max \{(1-2p_1)^2,(1-2p_2)^2\}.
\end{equation}

\begin{proof}[Proof]
	Consider the rate and distortion of (\ref{eq3}) are denoted by $R_{\text{e}}=0$ and $D_{\text{e}}=1$ if $(d_1^*,d_2^*)=(0.5,0.5)$. Calculation of the following fraction limit, which equals the slope of the tangent line to the sum-rate distortion curve, is desired;
	\begin{equation}
	\label{mumax}
	\lim_{(d_1,d_2)\to (0.5,0.5)} {D_{\text{e}}-D \over R-R_{\text{e}}} =
	\lim_{(d_1,d_2)\to (0.5,0.5)} {1+h_b(p*d)-h_b(p_1*d_1)-h_b(p_2*d_2) \over 1+h_b(p*d)-h_b(d_1)-h_b(d_2)}={0 \over 0}!
	\end{equation}
	
	By applying L'Hopital's rule and differentiation with respect to $d_1$, (\ref{mumax}) equals:
	\begin{align}
	\label{mumax1}
	\lim_{(d_1,d_2)\to (0.5,0.5)} {(1+m-2d_2-2md_1)(1-2p)h'_b(p*d)-(1-2p_1)h'_b(p_1*d_1)-m(1-2p_2)h'_b(p_2*d_2) \over (1+m-2d_2-2md_1)(1-2p) h'_b(p*d)-h'_b(d_1)-mh'_b(d_2)},
	\end{align}
	where $m={\partial d_2 \over \partial d_1}$. The above fraction is again ambiguous, therefore we need another differentiation with respect to $d_1$ from both the numerator and the denominator. After differentiation and letting $(d_1,d_2) \rightarrow (0.5,0.5)$,
	\begin{align}
	\label{mumax2}
	&\lim_{(d_1,d_2)\to (0.5,0.5)} {D_{\text{e}}-D \over R-R_{\text{e}}} =
	\lim_{(d_1,d_2)\to (0.5,0.5)} {0-(1-2p_1)^2h''_b(p_1*d_1)-m^2(1-2p_2)^2h''_b(p_2*d_2) \over 0-h''_b(d_1)-m^2h''_b(d_2)} \\ \nonumber
	&= {-(1-2p_1)^2h''_b(0.5)-m^2(1-2p_2)^2h''_b(0.5) \over -h''_b(0.5)-m^2h''_b(0.5)}
	= {(1-2p_1)^2+m^2(1-2p_2)^2 \over 1+m^2} \delequal g(m).
	\end{align}
	In the curve of the sum-rate distortion bound, the value of (\ref{mumax}) is maximized. Hence, the maximum of $g(m)$ is desirable.
	\begin{align}
	\label{mumax3}
	&g'(m)={\partial g(m) \over \partial m}= { \big(2m(1-2p_2)^2\big)(1+m^2)-2m\big((1-2p_1)^2+m^2(1-2p_2)^2\big)\over (1+m^2)^2}=0 \\ \nonumber &\Leftrightarrow \quad m=0 \quad \text{or} \quad m=\infty.
	\end{align}
	Therefore, the maximum value of the parameter $\mu$ is obtained from (\ref{mumax3}) as in (\ref{mu}).
\end{proof}
\textit{Corollary 3:} According to (\ref{mumax3}), the slope of the tangent line to the curve of location of the optimum points is $0$ if $p_1<p_2$, and it is $\infty$ if $p_1>p_2$. For the case $p_1=p_2$, both $0$ and $\infty$ are acceptable as the slope of the tangent line to the curve of location of the optimum points due to the continuity of the rate and the distortion functions.

In practical applications, parameters of the observation noises are small values. Hence, we investigate our problem with more details when at least one of the noise parameters is a very small value. Thus, we may assume without loss of generality that $p_1 \to 0$, then from continuity:
\begin{align}
\label{eq0}
R &\approx 1+h_b(p_2*d)-h_b(d_1)-h_b(d_2), \\ \nonumber 
D &\approx h_b(d_1)+h_b(p_2*d_2)-h_b(p_2*d) \approx 1-R+h_b(p_2*d_2)-h_b(d_2).
\end{align}
The behavior of low noise case is expressed by the following theorem.

\textit{Theorem 3:} If $p_1 \to 0$, then only two cases can occur: either (i) $d_2^*=0.5$, or (ii) $d_1^* \to 0$.
\begin{proof}[Proof]
	It is sufficient to prove that if $0 \le d_2^*<0.5$, then $d_1^* \to 0$. First, we shall declare that $h_b(q*x)-h_b(x)$ is a positive and a decreasing function in terms of $x$, where $0 \le x \le 0.5$ and $0 < q \le 0.5$. Therefore, it takes its maximum value in $x=0$ for any $q$. Now compute the objective function in the case $p_1 \to 0$. Due to (\ref{eq0}), we have:
	\begin{align}
	\label{eq00}
	F(\mu)=D+\mu R \approx 1+(\mu -1)R+h_b(p_2*d_2)-h_b(d_2).
	\end{align}
	Assume that $0 \le d_2^*=c<0.5$, where $c$ is a constant value. The latter optimization problem becomes as follows:
	\begin{equation}
	\begin{aligned}
	\label{opt5}
	& \underset{0 \le d_1 \le 0.5,\ d_2=c}{\text{min}}
	& & (\mu -1) \big(1+ h_b(p_2*d)-h_b(d_1)-h_b(d_2) \big)+h_b(p_2*d_2)-h_b(d_2) \equiv \\
	& \underset{0 \le d_1 \le 0.5}{\text{min}}
	& & (\mu -1) \big(h_b(q*d_1)-h_b(d_1) \big) \equiv  \underset{0 \le d_1 \le 0.5}{\text{max}}
	\big(h_b(q*d_1)-h_b(d_1) \big),
	\end{aligned}
	\end{equation}
	where $0 < q=p_2*c \le 0.5$. Obviously, solution of the above problem is $d_1^*=0$ and due to the approximations in our calculations, we shall have $d_1^* \to  0$.
\end{proof}

\begin{figure}[t]
	\centering
	\subfigure[{If $p_2=p_1$.}]{\label{s1}
		\includegraphics[width=2in,height=1.7in]{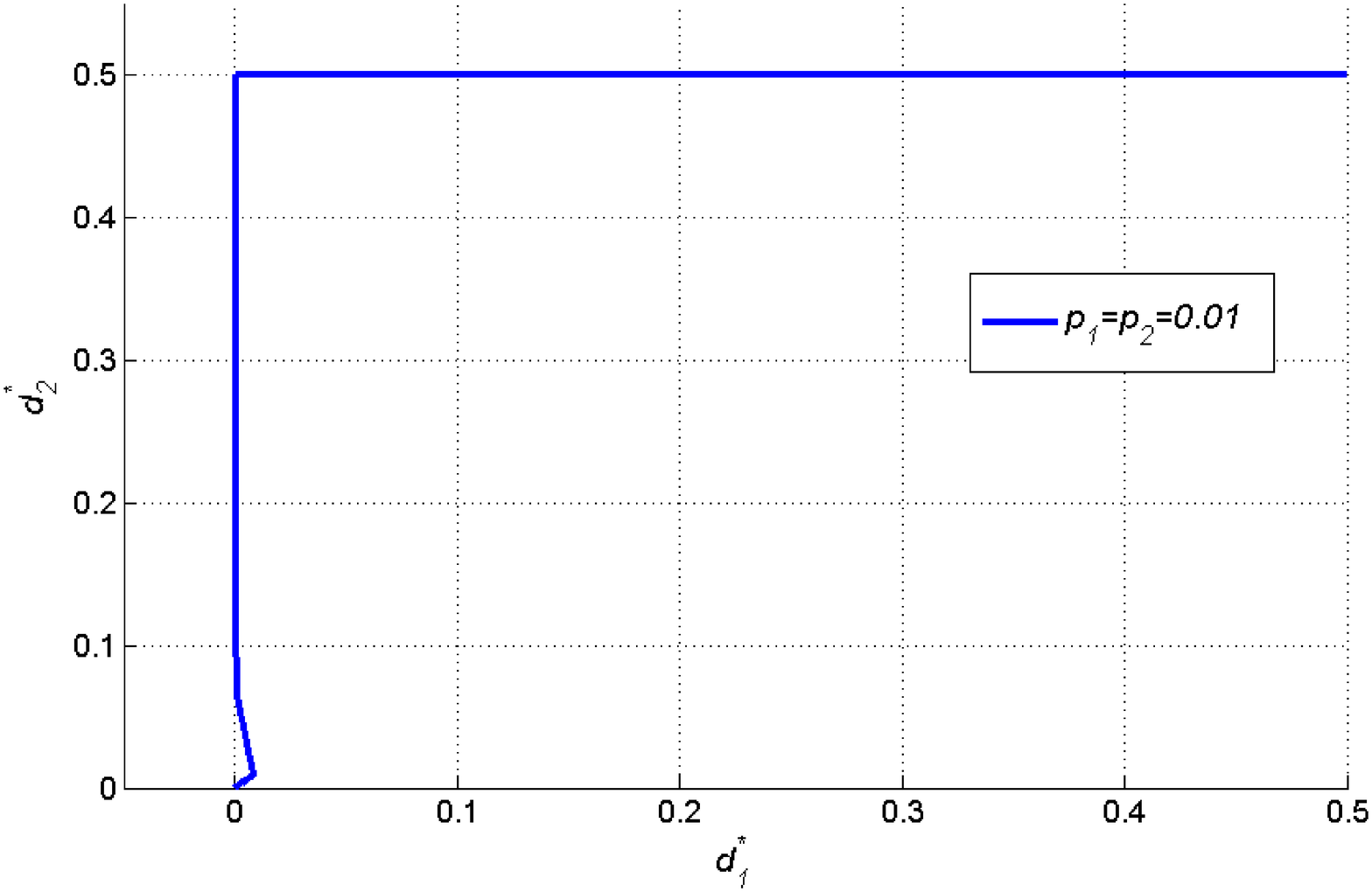}}
	\hspace{0.5mm}
	\subfigure[{If $p_2-p_1=0.01$.}]{\label{s2}
		\includegraphics[width=2in,height=1.7in]{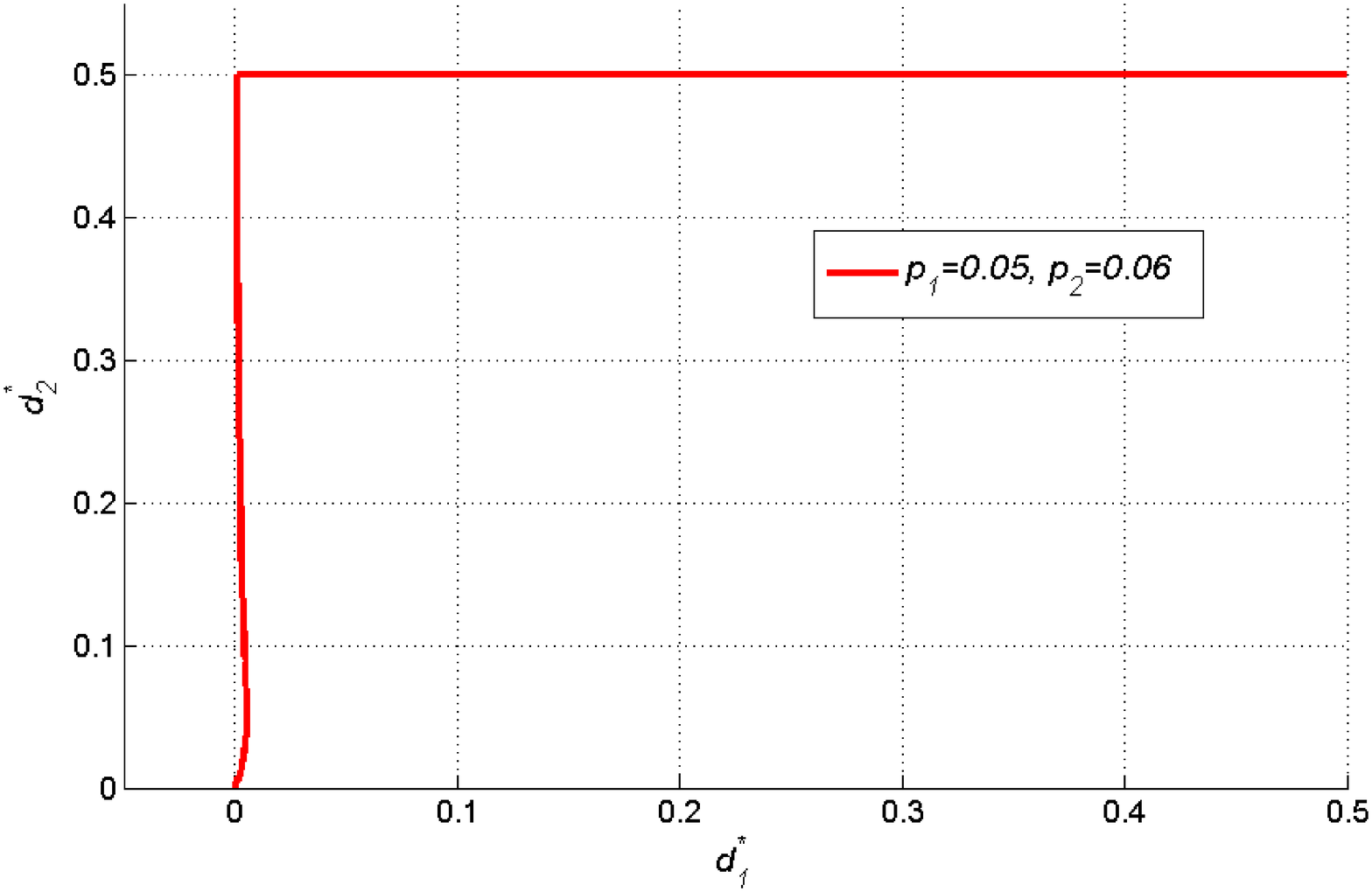}}
	\hspace{0.5mm}
	\subfigure[{If $p_2-p_1=0.05$.}]{\label{s3}
		\includegraphics[width=2in,height=1.7in]{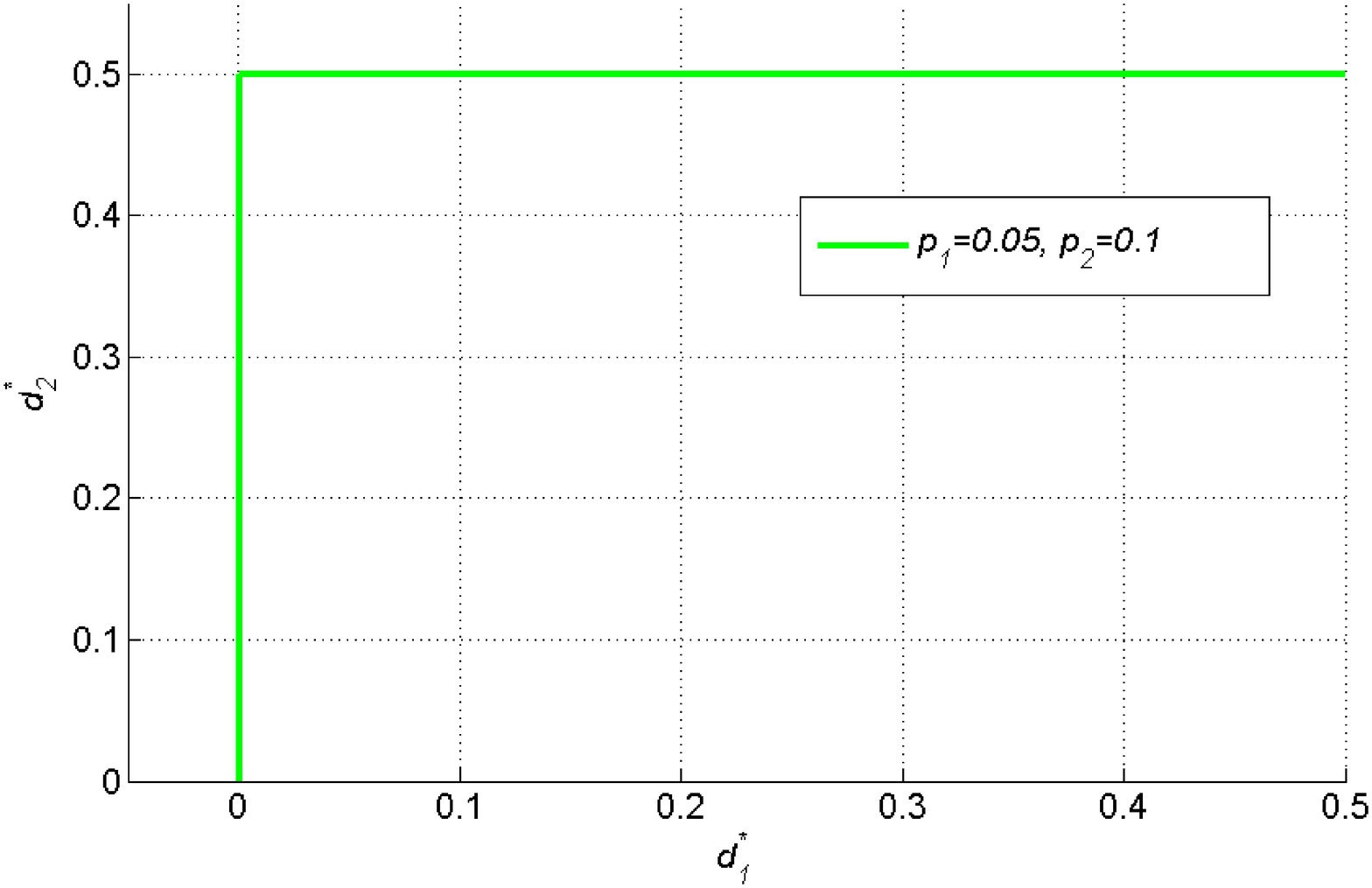}}
	\caption{Location of the optimum points $(d_1^*,d_2^*)$.}
	\label{s123}
\end{figure}
\textit{Corollary 4:} If $p_2$ is sufficiently larger than $p_1$, then $p=p_1*p_2 \approx p_2$ and the same situation of Theorem $3$ occurs. Similarly, if ${p_2 - p_1 \over p_2}= \alpha \to 1$, then $p=p_1*p_2=(1-\alpha)p_2+p_2-2(1-\alpha)p_1^2 \approx p_2$, and hence Theorem $3$ is used again, as depicted in the case of Fig. \ref{fig3}. In order to have more intuition to the result of Theorem $3$, some other cases are provided in the Fig. \ref{s123}.

\section{The Proposed Coding Scheme}
\begin{figure}[t]
	\begin{center}
		%\centering
		\includegraphics[width=3.5in,height=2.2in]{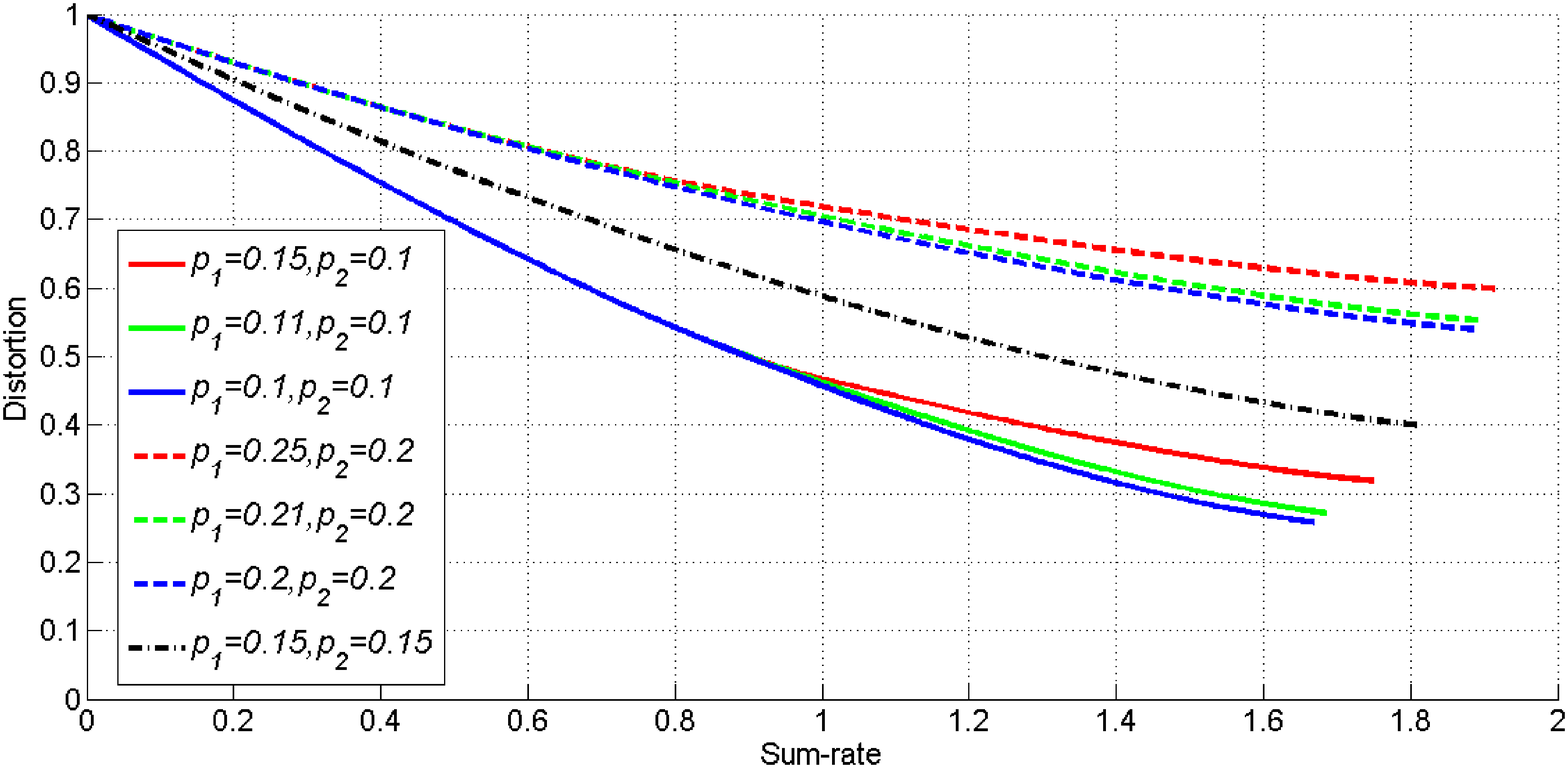}
		\vspace{-10pt}
		\caption{The sum-rate distortion function of the binary CEO for some noise parameters.}
		\label{RD}
	\end{center}
\end{figure}

In this section, a practical coding scheme is introduced to achieve the calculated rate-distortion bounds. In Fig. \ref{RD}, values of the theoretical bound of sum-rate versus distortion are displayed for several noise parameters to evaluate performance of the designed codes. The value of gap between the achieved point and the theoretical bound is employed as a performance criterion. The structure of the designed code significantly differs whether only one of the links or both of them are engaged in sending information. Obviously, when only one link sends information, our problem reduces to a point-to-point lossy source coding problem. Furthermore, for any $(d_1,d_2)$, there exists a particular achievable rate region which is characterized by the corner and intermediate points in its boundary. In the following, the proposed encoding and decoding schemes are separately illustrated for achieving the corner points and the intermediate points of the bound in the achievable rate region.

\subsection{Coding Scheme for the Corner Points}

\begin{figure}[t]
	\centering
	\subfigure[{The proposed encoding scheme based on compound structure.}]{\label{ENC1}
		\includegraphics[width=2.6in,height=1.5in]{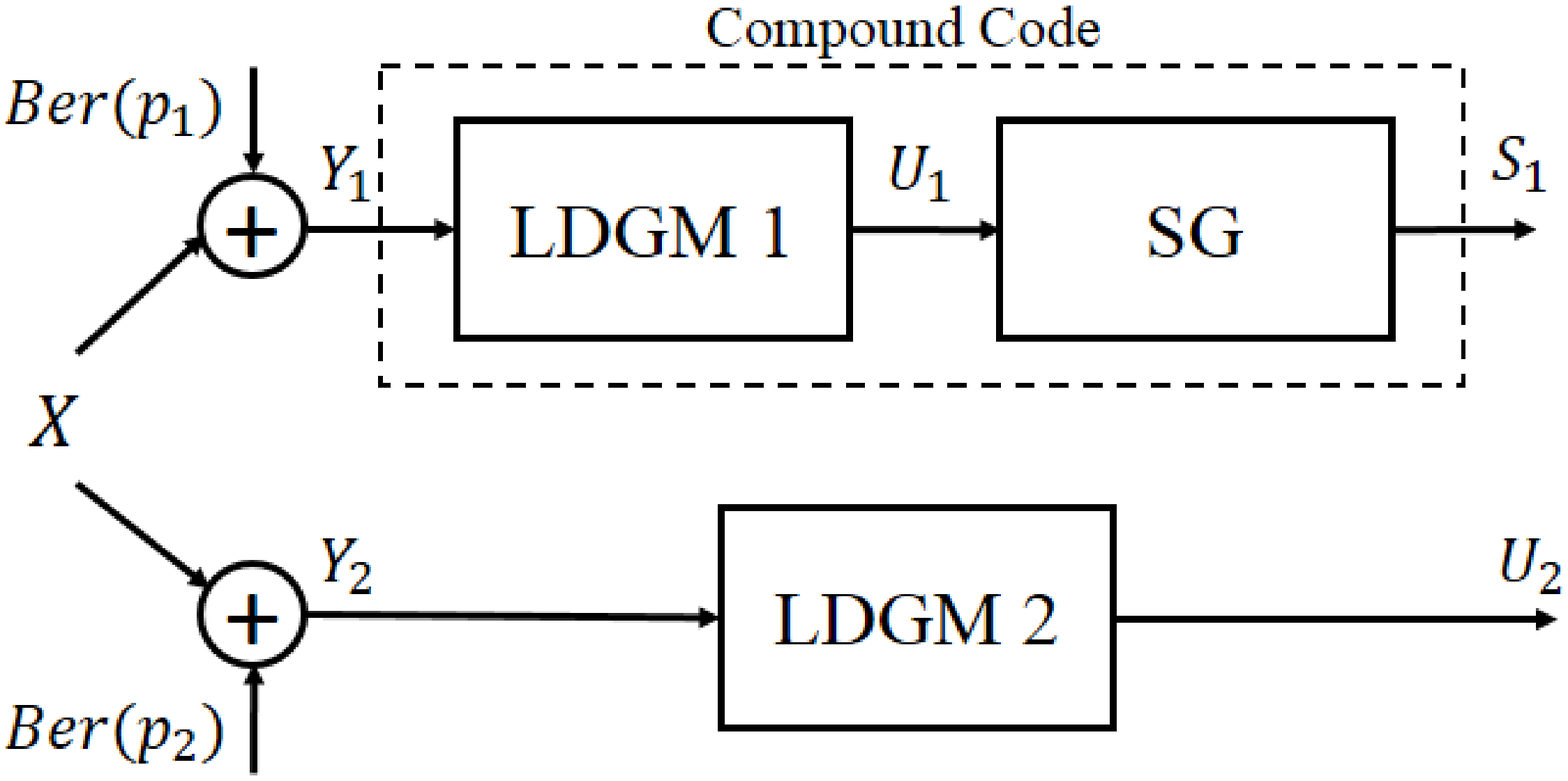}}
	\hspace{1mm}
	\subfigure[{Two-link joint decoding structure.}]{\label{jDEC1}
		\includegraphics[width=2.2in,height=1.5in]{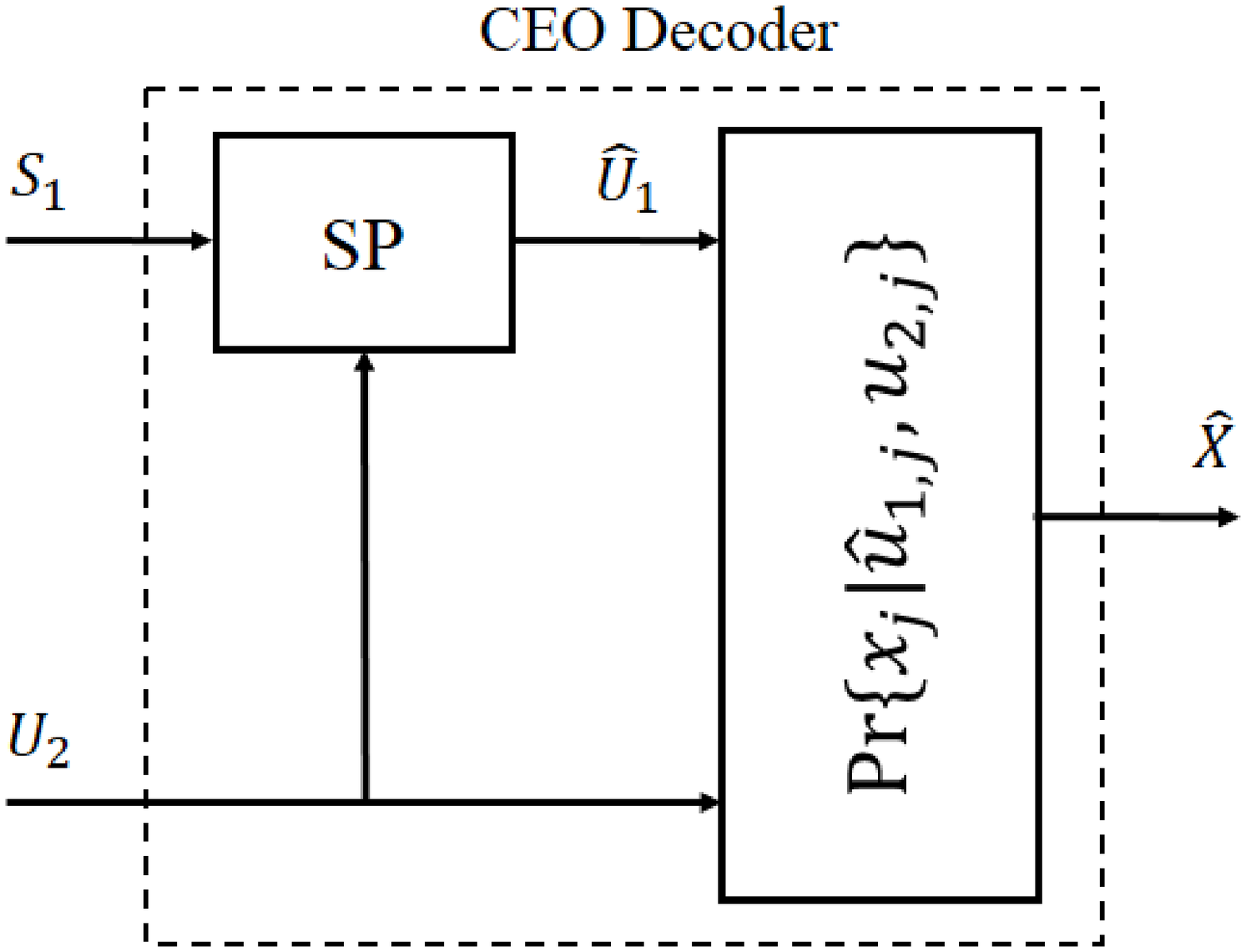}}
	\caption{The proposed coding scheme for achieving a corner point.}
	\label{scheme1}
\end{figure}

According to the exact rate-distortion bound (\ref{innerlog}), we have the following bound for the rate of $i$-th link and the sum-rate:
\begin{align}
\label{eq18}
&R_i \ge I({Y}_i ; {U}_i |  {U}_{3-i})=H( {U}_i |  {U}_{3-i})-H( {U}_i |  {Y}_i ,{U}_{3-i}) \\ \nonumber
&\stackrel{\text{(a)}}{=}H( {U}_i |  {U}_{3-i})-H( {U}_i |  {Y}_i )=h_b(d*p)-h_b(d_i),  \quad \text{for} \quad i=1,2, \\ \nonumber
& R_1+R_2 \ge I(Y_1,Y_2;U_1,U_2)\\ \nonumber
&=H(U_1,U_2)-H(U_1,U_2|Y_1,Y_2)\stackrel{\text{(a)}}{=}1+H(U_1|U_2)-H(U_1|Y_1)-H(U_2|U_2)\\ \nonumber
&=1+h_b(p*d)-h_b(d_1)-h_b(d_2),
\end{align}
where (a) follows from the Markov chain rule in the form of ${U}_1 \stackrel{d_1}\leftrightarrow {Y}_1 \stackrel{p_1} \leftrightarrow {X} \stackrel{p_2} \leftrightarrow {Y}_2 \stackrel{d_2} \leftrightarrow {U}_2$. Since a conventional rate-distortion quantizer can asymptotically achieve the compression rate $1-h_b(d_i)$ for distortion being assumed $d_i$, it is impossible to get close to (\ref{eq18}) by only using the rate-distortion quantizer. Hence, another lossless source encoder should be utilized after the conventional rate-distortion quantizer for achieving the rate (\ref{eq18}) in the $i$-th link. We use an LDGM quantizer concatenated with a Syndrome-Generator (SG), inspired by the ``quantize-and-bin'' idea in the context of information theory. On the other side, the dominant face of the rate region is a line segment connecting two end points $(R'_1,R'_2)$ and $(R''_1,R''_2)$, where
\begin{align}
\label{RR}
(R'_1,R'_2)=\big(h_b(p*d)-h_b(d_1),1-h_b(d_2)\big),
\end{align}
and
\begin{align}
\label{RR1}
(R''_1,R''_2)=\big(1-h_b(d_1),h_b(p*d)-h_b(d_2)\big).
\end{align} 
We consider a coding scheme for achieving $(R'_1,R'_2)$. A similar method can be applied for achieving $(R''_1,R''_2)$. Encoder $1$ quantizes $y_1^n$ to $u_1^n$ using an LDGM code of rate $R_{1,1}={m_1 \over n}$, then it computes the syndrome $s_1^{k_1}=u_1^n H_1^T$, where $H_1$ is the parity-check matrix of an LDPC code of rate $R_{1,2}={m_1-k_1 \over n}$. We do this process of quantize and bin by employing a compound LDGM-LDPC code. It is notable that the total length of the obtained syndrome equals $n-m_1+k_1$, where its first $n-m_1$ bits are zero because the LDPC code is nested in the LDGM code \cite{NAA17}. Hence, only $k_1$ non-zero bits are sent to the decoder and the total rate is $R_1=R_{1,1}-R_{1,2}={k_1 \over n}$. Encoder $2$ quantizes $y_2^n$ to $u_2^n$ using an LDGM code of rate $R_{2,1}={m_2 \over n}$, then $u_2^n$ is sent to the decoder. In the second link, the total rate is $R_2={m_2 \over n}$. The block diagram of the proposed scheme is shown in Fig. \ref{scheme1}.

At the decoder side, the syndrome $s_1^{k_1}$ with $u_2^n$ as a side information are used to decode $u_1^n$, denoted by $\hat u_1^n$, by applying a SP algorithm. Finally, at the decoder, calculation of the soft estimation $\hat x_j=\Pr\{x_j | \hat u_{1,j}, u_{2,j}\}$ completes the decoding process.
Here $\Pr\{x_j | \hat u_{1,j}, u_{2,j}\} \delequal p_{X|U_1,U_2}(x_j|\hat u_{1,j}, u_{2,j})$, and the conditional distribution $p_{X|U_1,U_2}$ can be deduced from the joint distribution in (\ref{jeq}) once $d_1$ and $d_2$ are given (assuming $Q$ is a constant).

%Previously, we have used the compound LDGM-LDPC codes for achieving theoretical bound of the binary Wyner-Ziv problem in \cite{NAA17}. %Structure of the compound LDGM-LDPC code in the $i$-th link is shown in Fig. \ref{comp}, and the details are provided in \cite{NAA17}.
%\begin{figure}[t]
%	\begin{center}
%		%\centering
%		\includegraphics[width=3.3in,height=2.2in]{comp.eps}
%		\vspace{-20pt}
%		\caption{The Tanner graph of the compound LDGM-LDPC codes.}
%		\label{comp}
%	\end{center}
%\end{figure}
A compound LDGM-LDPC code includes nested LDGM and LDPC codes with the following parity-check matrices:
\begin{equation}
\label{matr}
H_{\text{LDPC}}=
\begin{bmatrix}
H_{\text{LDGM}} \\
\Delta H \\
\end{bmatrix},
\end{equation}
where $H_{\text{LDPC}}$ and $H_{\text{LDGM}}$ are, respectively, parity-check matrices of the LDPC and LDGM codes. Let assume their sizes are $(n-m+k) \times n$ and $(n-m) \times n$, respectively. We have used the compound LDGM-LDPC structure to achieve theoretical bound of the Wyner-Ziv problem in \cite{NAA17}. We denote the mentioned compound code by $\mathcal{C}_{H_{\text{LDPC}}}(n,m,k)$.

For achieving a corner point, we employ a compound code $\mathcal{C}_{H^{(1)}_{\text{LDPC}}}(n,m_1,k_1)$ in the first link, and a single LDGM code with the generator matrix $G^{(2)}$ of size $m_2 \times n$ in the second link. The observation $y_1^n$ is quantized to an LDGM codeword $u_1^n$ by applying the BiP algorithm with the generator matrix $G^{(1)}$. Hence, $u_1^n H_{\text{\text{LDGM}}}^{(1)}=\underbrace{[0 \ \cdots \ 0]}_{n-m_1}$. Next, the syndrome $u_1^n H_{\text{\text{LDPC}}}^{(1)}=[\underbrace{0 \ \cdots \ 0}_{n-m_1} ,\ \underbrace{u_1^n \Delta H^T}_{s_1^{k_1}}]$ is calculated and only $s_1^{k_1}$ is sent to the CEO decoder.
 
\subsection{Coding Scheme for the Intermediate Points}

Consider the following intermediate point located in the dominant face of the achievable rate region,
\begin{equation}
\label{intermed}
(R_1^*,R_2^*)=\big(h_b(p*d)-h_b(d_1)+\delta,1-h_b(d_2)-\delta\big), 
\end{equation}
where $0 < \delta <1-h_b(p*d)$. Obviously, $R_1^* \le 1-h_b(d_1)$ and $R_2^* \le 1-h_b(d_2)$. Therefore, a lossless compression is needed in each link. In the $i$-th link, we use a compound code $\mathcal{C}_{H^{(i)}_{\text{LDPC}}}(n,m_i,k_i)$, for $i=1,2$. First step of encoding includes quantizing the observations $y_i^n$ to $u_i^n$ by using the BiP algorithm on the LDGM codes associated with the parity-check matrices $H^{(i)}_{\text{LDGM}}$. In the second step, the syndromes $u_i^n H_{\text{\text{LDPC}}}^{(i)}=[\underbrace{0 \ \cdots \ 0}_{n-m_i} ,\ \underbrace{u_i^n \Delta {H^{(i)}}^T}_{s_i^{k_i}}]$ are calculated, then only $s_1^{k_1}$ and $s_2^{k_2}$ are sent to the CEO decoder.

In the decoder, we propose a Joint Sum-Product (JSP) algorithm which is a modified version of the SP algorithm. In this algorithm, the received syndromes $s_1^{k_1}$ and $s_2^{k_2}$ are respectively located in the check nodes of the LDPC codes with parity-check matrices $H^{(1)}_{\text{LDPC}}$ and $H^{(2)}_{\text{LDPC}}$. The JSP includes $r$ rounds and each round includes $l$ iterations. At the starting point of the JSP, initial LLRs in the variable nodes are set based on a random side information in each SP. At the end of each round, which includes update equations in the check and variable nodes, the bit values of the variable nodes are calculated according to the decision rule of the SP algorithm, where it maps the non-negative LLRs to bit $0$ and the negative LLRs to bit $1$. In the next round, these updated bit values in the variable nodes are used as a new side information for calculating new initial LLR values. Finally, after $r$ rounds, $\hat u_1^n$ and $\hat u_2^n$ are decoded based on the decision rule of the SP algorithm in the variable nodes of the LDPC codes. An EXIT chart analysis is presented in \cite{KSA17} for a similar JSP decoder which shows the capacity approaching property with two parallel and collaborative SP decoders. Similar to the decoding scheme of the corner points, the soft estimation $\hat x_j=\Pr\{x_j | \hat u_{1,j}, \hat u_{2,j}\}$ accomplishes the decoding process.

\begin{figure}[t]
	\centering
	\subfigure[{The proposed encoding scheme based on compound structures.}]{\label{ENC}
		\includegraphics[width=2.6in,height=1.5in]{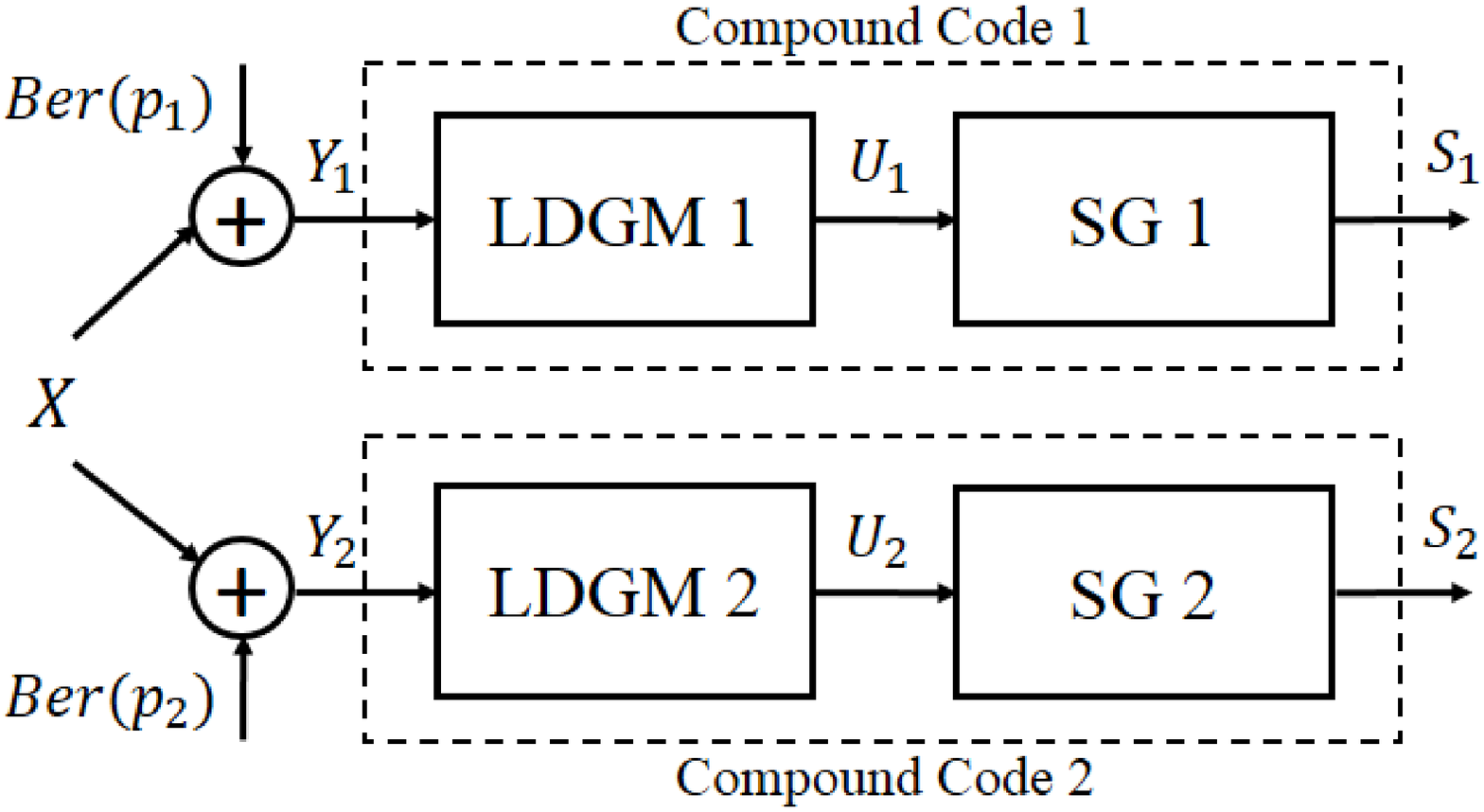}}
	\hspace{1mm}
	\subfigure[{Two-link joint decoding structure.}]{\label{jDEC}
		\includegraphics[width=2.2in,height=1.6in]{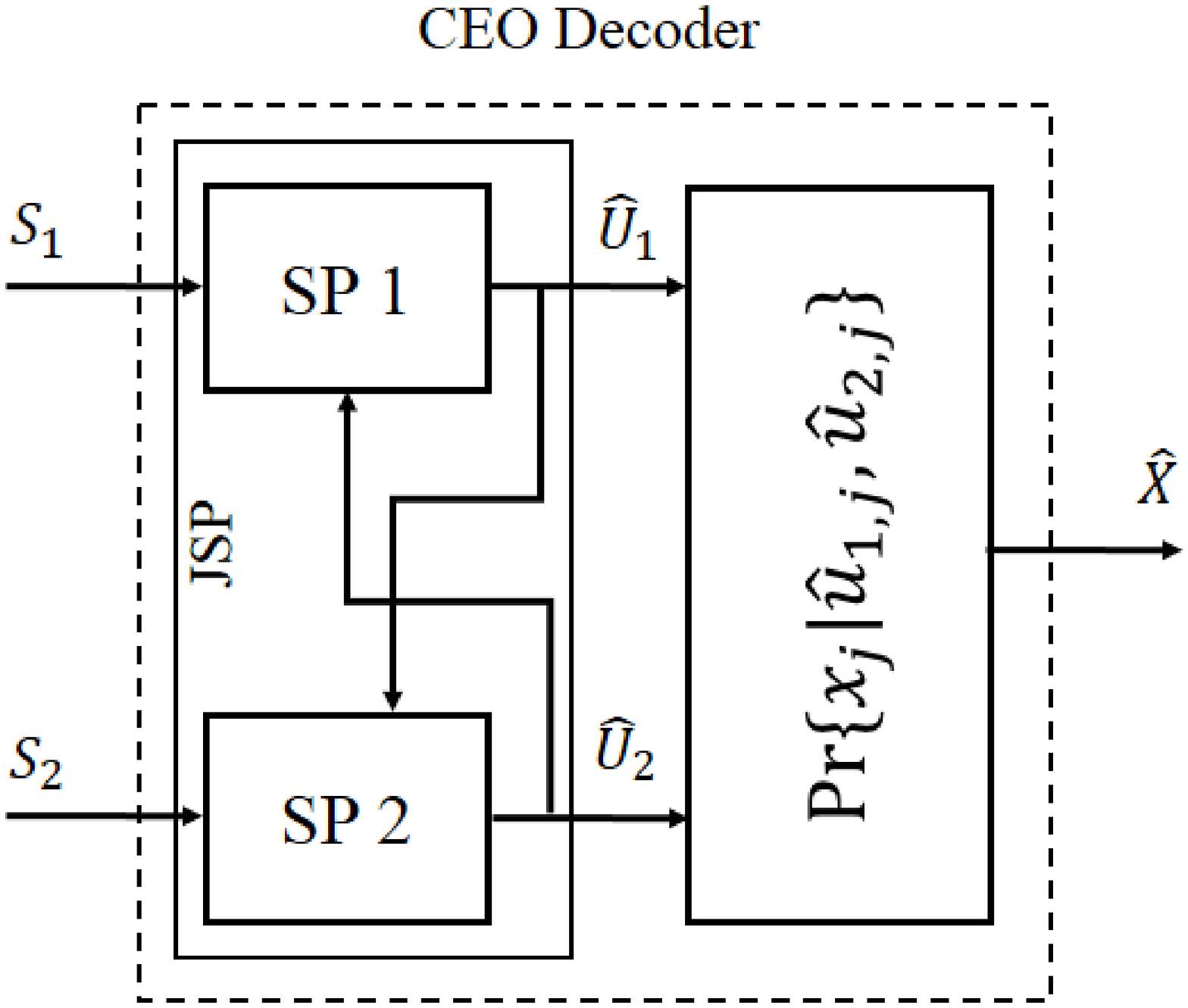}}
	\caption{The proposed coding scheme for achieving intermediate points.}
	\label{scheme}
\end{figure}

In the joint decoding scheme, the received sequences ${s}_1^n$ and ${s}_2^n$ are simultaneously decoded. If we look at this situation like two point-to-point lossy source coding problems, then we have to recover the noisy observations ${y}_1^n$ and ${y}_2^n$, each of which is received with compression rates $R_1$ and $R_2$ and acceptable distortions, respectively. As a case, let these distortions be $d_1$ and $d_2$, respectively. A major part of distortions $d_1$ and $d_2$ arises from the LDGM quantization and a negligible part is from the syndrome decoding. Let assume the LDPC code rate in the $i$-link is denoted by $R_{i,2}$, for $i=1,2$. Furthermore, consider the associated distortion of each link, i.e., BER of the syndrome-decoding part, is denoted by $d_{i,2}$, for $i=1,2$. Using the compound LDGM-LDPC structure, the total rate and the distortion in each link are as follows:
\begin{align}
\label{trd}
d_i=d_{i,1}*d_{i,2} \approx d_{i,1}, \ R_i=R_{i,1}-R_{i,2}, \quad \text{for} \quad i=1,2.
\end{align}
After reconstruction of the observations with distortions $d_1$ and $d_2$, denoted by $\hat{{u}}_1^n$ and $\hat{{u}}_2^n$, the soft reconstruction of the original binary source $\hat {x}^n$ is estimated.
\subsection{A Practical Analysis for the Proposed Coding Scheme}
Some coding parameters are affected by the information theoretical limits, that should be considered in the code design procedure. In the following notations, any $\epsilon$ denotes a sufficiently small positive value. In the coding scheme for a corner point, the relation between the rate-distortion and the block lengths of employed LDGM and LDPC codes are as follows:
\begin{align}
\label{eq19}
R_{1,1}&={m_1 \over n}=1-h_b(d_{1,1})+\epsilon_{1,1}, \ R_{1,2}={m_1-k_{1} \over n}=1-h_b(d_{1,1}*d_{2,1}*p_1*p_2)-\epsilon_{1,2}, \\ \nonumber
R_{2,1}&={m_2 \over n}=1-h_b(d_{2,1})+\epsilon_{2,1}, \ R_{2,2}=0.
\end{align}
From (\ref{trd}) and (\ref{eq19}), it is simply concluded that:
\begin{align}
\label{eq21}
R_1&=R_{1,1}-R_{1,2}={k_{1} \over n} \\ \nonumber &=h_b(d_{1,1}*d_{2,1}*p_1*p_2)-h_b(d_{1,1})+\underbrace{\epsilon_{1,1}+\epsilon_{1,2}}_{\epsilon_1}
\stackrel{\text{(b)}}{\approx} h_b(d*p)-h_b(d_{1})+{\epsilon_1}, \\ \nonumber
R_2&=R_{2,1}-R_{2,2}={k_{2} \over n}=1-h_b(d_{2,1})+\epsilon_{2,1} \stackrel{\text{(b)}}{\approx} 1-h_b(d_{2})+\epsilon_{2,1},
\end{align}
where (b) follows from the continuity of the function $h_b(x)$. The above approximation expresses that achieving the rate bound (\ref{eq18}) is possible by using our proposed method in each link. 
In the decoding side, the SP algorithm of Fig. \ref{jDEC1} uses the LDPC code of rate $R_{1,2}$, that is smaller than the capacity of the virtual channel between the side information $U_2$ and the target sequence $U_1$ \footnote{This capacity equals $1-h_b(p*d)$.}. Therefore, $U_1$ is decoded with a low BER, i.e., $d_{1,2} \approx 0$, by using a good channel decoder. Clearly, in this case $d_{2,2}=0$ and $d_2=d_{2,1}$.

In the coding scheme for an intermediate point (\ref{intermed}), the relation between the rate-distortion and the block lengths of each employed LDGM and LDPC codes are as follows:
\begin{align}
\label{eq19i}
R_{1,1}&={m_1 \over n}=1-h_b(d_{1,1})+\epsilon_{1,1}, \ R_{1,2}={m_1-k_{1} \over n}=1-h_b(d_{1,1}*d_{2,1}*p_1*p_2)- \delta -\epsilon_{1,2}, \\ \nonumber
R_{2,1}&={m_2 \over n}=1-h_b(d_{2,1})+\epsilon_{2,1}, \ R_{2,2}={m_2-k_{2} \over n}=\delta-\epsilon_{2,2}.
\end{align}
From (\ref{trd}) and (\ref{eq19i}), it is simply concluded that:
\begin{align}
\label{eq21i}
R_1&=R_{1,1}-R_{1,2}={k_{1} \over n} \\ \nonumber &=h_b(d_{1,1}*d_{2,1}*p_1*p_2)-h_b(d_{1,1})+ \delta +\underbrace{\epsilon_{1,1}+\epsilon_{1,2}}_{\epsilon_1}
\stackrel{\text{(b)}}{\approx} h_b(d*p)-h_b(d_{1})+\delta+{\epsilon_1}, \\ \nonumber
R_2&=R_{2,1}-R_{2,2}={k_{2} \over n}=1-h_b(d_{2,1})-\delta+\underbrace{\epsilon_{2,1}+\epsilon_{2,2}}_{\epsilon_2} \stackrel{\text{(b)}}{\approx} 1-h_b(d_{2})-\delta+\epsilon_{2},
\end{align}
where (b) follows from the continuity of the function $h_b(x)$. The above approximation expresses that achieving the rate bound (\ref{eq18}), for an intermediate point (\ref{intermed}), is possible by utilizing the proposed method. In the decoding side, the JSP algorithm of Fig. \ref{jDEC} uses the LDPC codes of rates $R_{1,2}$ and $R_{2,2}$. From (\ref{eq19i}) and $0<\delta<1-h_b(p*d)$,
\begin{align}
\label{rLD}
R_{1,2}&\approx 1-h_b(d*p)- \delta -\epsilon_{1,2}<1-h_b(d*p), \\ \nonumber
R_{2,2}&=\delta-\epsilon_{2,2}<1-h_b(d*p).
\end{align}
Therefore, the rates of LDPC codes in the SP$1$ and the SP$2$ algorithms are smaller than the capacity of the virtual channel between the side information $U_1$ and $U_2$. This implies that the SP algorithms can decode $U_1$ and $U_2$ with low BERs, i.e., $d_{i,2} \approx 0$ for $i=1,2$, for sufficiently large $n$, $r$, and $l$.

For the empirical distortion $D_{\text{em}}$ in (\ref{eqlogg}) with $\hat x_j(x_j)=\Pr\{x_j|u_{1,j},u_{2,j}\}$, we have
\begin{align}
\label{eq25}
D_{\text{em}}={1 \over n} \sum_{j=1}^n \log [{1 \over \Pr\{x_j|u_{1,j},u_{2,j}\}}]=\sum_{x,u_1,u_2} \Pr_{\text{em}}\{x,u_1,u_2\} \log [{1 \over \Pr\{x|u_1,u_2\}}],
\end{align}
where $\Pr_{\text{em}}\{x,u_1,u_2\}$ is the empirical distribution induced by $(x^n,u_1^n,u_2^n)$. The theoretical distortion bound is given by $D_{\text{th}}=H(X|U_1,U_2)$. Clearly, we have $D_{\text{em}} \approx D_{\text{th}}$ if $\Pr_{\text{em}}\{x,u_1,u_2\}$ is close to $\Pr\{x,u_1,u_2\}$.

\section{Results and Discussions}

In this section, some numerical results are given for indicating the rate-distortion performance of the proposed coding scheme at different regions. For all of the LDPC codes, the optimized degree distributions over the BSC are employed \footnote{These degree distributions are available in \cite{SC07} for some rates.}. However, the check-regular and variable-Poisson LDGM codes nested with the LDPC codes are designed similar to the code design method in \cite{NAA17}. In order to achieve some target optimum crossover probability pairs $(d_1^*,d_2^*)$ by practical coding methods, we have applied our proposed coding scheme with the lengths of $n=10^4, \ 10^5$ for various cases of the noise parameters including $(p_1,p_2)=(0.15,0.15)$ and $(0.29,0.3)$. We have also implemented our proposed coding scheme for two low-noise cases $(0.01,0.01)$ and $(0.05,0.1)$.

In the BiP algorithm, the parameters $t=0.8$, $\gamma_i \approx 2 R_{i,1}=2{m_i \over n}$ are selected for $i=1,2$. Maximum number of iterations in each round of this algorithm is set to be $25$. In the SP algorithm, maximum number of iterations is set to be $100$. Also, in the JSP algorithm, $r=15$ and $l=40$. All of the reported values for the empirical distortions are averaged over $50$ runs. Parameters of the employed codes and their results are presented in Table \ref{t1}. The corner and the intermediate points are indicated by symbols C and I in the second column of the table, respectively. The gap value is equal to difference between the empirical distortion $D_{\text{em}}$ and the theoretical distortion $D_{\text{th}}$ and it evaluates performance of the designed codes.

\begin{table}[t]
%\centering
\caption{\footnotesize{NUMERICAL RESULTS OF THE PROPOSED ENCODING AND DECODING METHODS.}}
\label{t1}
\centering
\vspace{-15pt}
\begin{center}
	\scalebox{.8}{
		%\begin{tabular} {| l | l | l | l | l | l | l | l | l | l | l | l | l |}
		\begin{tabular} {| c | c | c | c | c | c | c | c | c | c | c | c |}
			\hline
			$(p_1,p_2)$ &Region-C/I& $n$ & $m_1,m_2$ & $k_1,k_2$ & $d_{1,1},d_1$ & $d_{2,1},d_2$ &  $\mu$ & $R_{\text{th}}$ &  $D_{\text{th}}$ & $D_{\text{em}}$ &  $\text{Gap}$ \\
			\hline
			$(0.15,0.15)$ & $1$-I & $10^4$ & $9200,9200$ & $8500,8500$& $0.0144,0.0175$ & $0.0144,0.0175$ & $0.168$& $1.6722$ & $0.4204$ & $0.4617$ & $0.0413$ \\
			
			$(0.15,0.15)$ & $1$-I & $10^4$ & $5400,5400$ & $5100,5100$& $0.1028,0.1071$ & $0.1028,0.1071$ & $0.326$& $0.9898$ & $0.5925$ & $0.645$ & $0.0525$ \\
			
			$(0.15,0.15)$ & $1$-C & $10^4$ & $5400,5400$ & $4700,5400$& $0.1028,0.1076$ & $0.1028,0.1028$ & $0.326$& $0.9898$ & $0.5925$ & $0.6355$ & $0.043$ \\
			
			$(0.15,0.15)$ & $2$-I & $10^4$ & $5400,1300$ & $5300,1200$& $0.1028,0.1066$ & $0.3055,0.3078$ & $0.3854$& $0.6319$ & $0.7206$ & $0.7766$ & $0.056$ \\
			
			$(0.15,0.15)$ & $3$-C & $10^4$ & $5400,-$ & $5400,-$& $0.1028,0.1028$ & $0.5,0.5$  & $0.4043$& $0.531$ & $0.7601$ & $0.7955$ & $0.0354$ \\
			
			$(0.15,0.15)$ & $3$-C & $10^4$ & $1300,-$ & $1300,-$& $0.3055,0.3055$ & $0.5,0.5$& $0.4532$& $0.1187$ & $0.9427$ & $0.9835$ & $0.0408$ \\
			
			$(0.15,0.15)$ & $1$-I & $10^5$ & $92000,92000$ & $85000,85000$& $0.012,0.015$ & $0.012,0.015$ & $0.168$& $1.6722$ & $0.4204$ & $0.4451$ & $0.0247$ \\
			
			$(0.15,0.15)$ & $1$-I & $10^5$ & $54000,54000$ & $51000,51000$& $0.1003,0.1043$ & $0.1003,0.1043$ & $0.326$& $0.9898$ & $0.5925$ & $0.6203$ & $0.0278$ \\
			
			$(0.15,0.15)$ & $1$-C & $10^5$ & $54000,54000$ & $47000,54000$& $0.1003,0.105$ & $0.1003,0.1003$ & $0.326$& $0.9898$ & $0.5925$ & $0.6184$ & $0.0259$ \\
			
			$(0.15,0.15)$ & $2$-I & $10^5$ & $54000,13000$ & $53000,12000$& $0.1003,0.1037$ & $0.3018,0.304$ & $0.3854$& $0.6319$ & $0.7206$ & $0.7494$ & $0.0288$ \\
			
			$(0.15,0.15)$ & $3$-C & $10^5$ & $54000,-$ & $54000,-$&  $0.1003,0.1003$ & $0.5,0.5$ & $0.4043$& $0.531$ & $0.7601$ & $0.7826$ & $0.0225$ \\
			
			$(0.15,0.15)$ & $3$-C & $10^5$ & $13000,-$ & $13000,-$& $0.3018,0.3018$ & $0.5,0.5$ & $0.4532$& $0.1187$ & $0.9427$ & $0.9707$ & $0.028$ \\
			\hline
			$(0.29,0.3)$ & $1$-I & $10^4$ & $5420,2920$ & $5400,2900$ & $0.1025,0.1066$ & $0.2044,0.208$ & $0.1283$& $0.8044$ & $0.8797$ & $0.9423$ & $0.0626$ \\
			
			$(0.29,0.3)$ & $2$-C & $10^4$ & $2900,-$ & $2900,-$&  $0.2046,0.2046$ & $0.5,0.5$ &$0.157$& $0.278$ & $0.9537$ & $0.9942$ & $0.0405$ \\
			
			$(0.29,0.3)$ & $1$-I & $10^5$ & $54200,29200$ & $54000,29000$& $0.1,0.1038$ & $0.202,0.2052$ & $0.1283$& $0.8044$ & $0.8797$ & $0.9127$ & $0.033$ \\
			
			$(0.29,0.3)$ & $2$-C & $10^5$ & $29000,-$ & $29000,-$&  $0.2025,0.2025$ & $0.5,0.5$ & $0.157$& $0.278$ & $0.9537$ & $0.9786$ & $0.0249$ \\
			\hline
			$(0.01,0.01)$ & $1$-I & $10^4$ & $9900,9900$ & $6000,6000$& $0.0032,0.0053$ & $0.0032,0.0053$ & $0.4$& $1.1161$ & $0.0268$ & $0.0372$ & $0.0104$ \\
			
			$(0.01,0.01)$ & $1$-I & $10^5$ & $99000,99000$ & $60000,60000$&  $0.0028,0.0047$ & $0.0028,0.0047$ & $0.4$& $1.1161$ & $0.0268$ & $0.0329$ & $0.0061$ \\
			\hline
			$(0.05,0.1)$ & $1$-C & $10^4$ & $10000,400$ & $10000,300$&  $0,0$ & $0.405,0.4075$ & $0.253$& $1.014$ & $0.2829$ & $0.31$ & $0.0271$ \\
			
			$(0.05,0.1)$ & $1$-C & $10^5$ & $100000,4000$ & $100000,3000$& $0,0$ & $0.4026,0.4043$ & $0.253$& $1.014$ & $0.2829$ & $0.3011$ & $0.0182$ \\
			\hline
	\end{tabular}}
\end{center}
\end{table}

\begin{figure}[t]
	\centering
	\subfigure[{$p_1=p_2=0.15.$}]{\label{RD1}
		\includegraphics[width=3.1in,height=2in]{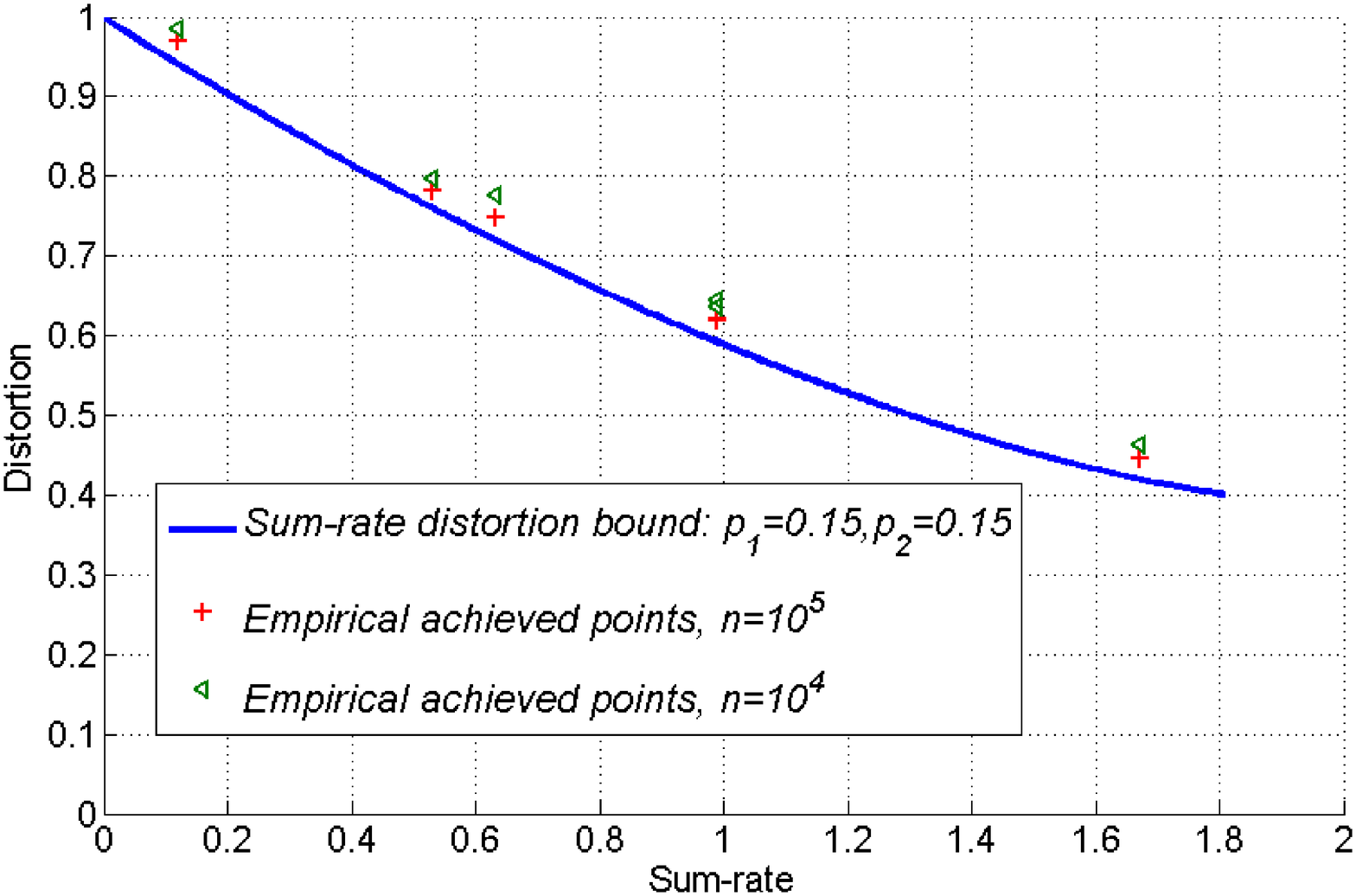}}
	\hspace{1mm}
	\subfigure[{$(p_1,p_2)=(0.29,0.3).$}]{\label{RD2}
		\includegraphics[width=3.1in,height=2in]{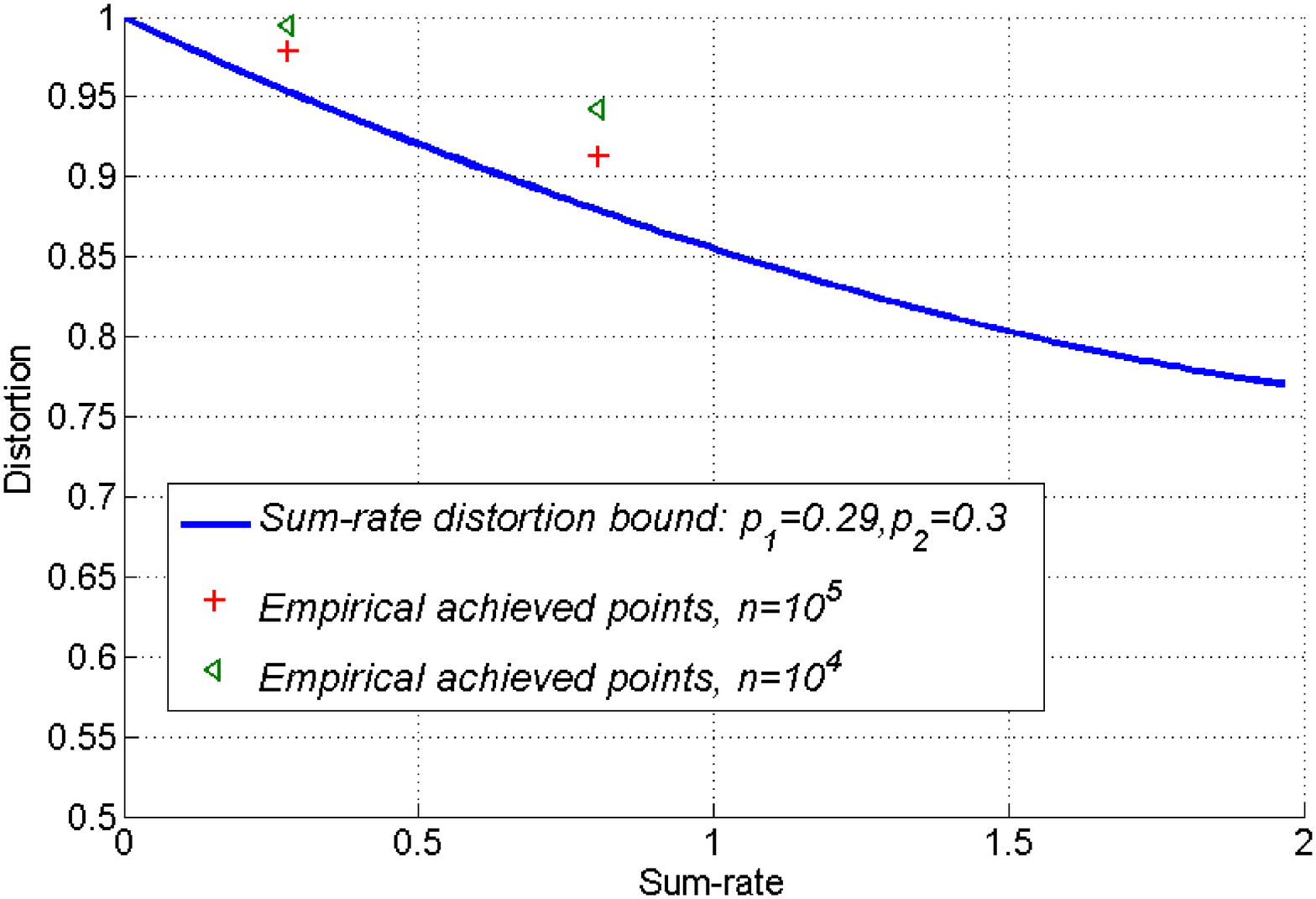}}
	\vspace{-15pt}
	\caption{Performance of the sum-rate versus distortion for the proposed coding scheme.}
	\label{performance}
\end{figure}

For the case of equal noises $(p_1,p_2)=(0.15,0.15)$, the gap values is about $0.03$ to $0.06$ for the block length $10^4$, as indicated in Table. \ref{t1}. It is obvious that by increasing the target distortions, the gap values increase. As the block length is set to $10^5$, the gap value decreases in the range between $0.02$ to $0.03$. Performance of the sum-rate versus distortion is depicted in Fig. \ref{RD1} for the proposed coding scheme. Simulation results confirm that performance of the sum-rate in terms of distortion is very close to the theoretical bounds for the empirically achieved points. The theoretical bounds are asymptotically achievable by employing the proposed coding scheme as well. For the case of unequal noise parameters $(p_1,p_2)=(0.29,0.3)$, the achieved gaps are slightly more than that one for the case of equal noise parameters with approximately the same target distortions and block length. This observation expresses that increasing noise parameters $p_1$ and $p_2$ causes a higher gap values in distortion. Similarly, by increasing the block length to $10^5$, the gap value is decreased by about less than half of the result for block length $10^4$, as mentioned in Table. \ref{t1}. Performance of the sum-rate in terms of distortion for the proposed coding scheme is depicted in Fig. \ref{RD2}.

In the low-noise case, as it is seen in the Table. \ref{t1}, the achieved gap depends on the value of the target distortions, similar to the prior cases. The achieved gap with our coding scheme is about $0.006$ to $0.011$ for the case $(p_1,p_2)=(0.01,0.01)$, and it is about $0.018$ to $0.028$ for the case $(p_1,p_2)=(0.05,0.1)$. Generally, the gap value for the corner points is smaller than that of the intermediate points. This gap can be reduced by increasing $r$ and $l$.

\section{Conclusion}

In this paper, we investigated the two-link binary CEO problem under the log-loss from both information theory and coding theory points of view. Under this criterion, the exact achievable rate-distortion region of the two-link binary CEO problem was calculated. Furthermore, optimal test-channel models were obtained for the encoders. By assuming the BSC test-channel model, we found optimal values of the crossover probabilities and then they are analyzed in a high-resolution regime, asymptotically. In the coding part, we proposed a practical coding scheme based on graph-based codes and message-passing algorithms. In the encoding side, a binary quantization and a syndrome-generation are utilized in each link for construction of the lossy compressed sequences. This was realized by the compound LDGM-LDPC codes. In the decoding part, the SP algorithms based on the optimized LDPC codes for the BSCs were employed at the first step. Then, a soft decoder calculated the final reconstruction value in the form of a probability distribution. Our experimental simulation results confirmed that the proposed coding scheme asymptotically achieves the theoretical bound of the two-link binary CEO problem under the log-loss. For a finite block length, there remains a slight gap between the rate-distortion of the proposed scheme and the associated theoretical bound which is a useful measure for comparison of the different coding methods. Future study and researches may extend the contents of this paper to the multi-link binary CEO problem and also to the multi-terminal lossy source coding problem.

%\appendices
%\section{Proof of the}

%\ifCLASSOPTIONcaptionsoff
%\newpage
%\fi

\bibliographystyle{IEEEtran}
{%\footnotesize
	\bibliography{refs}}

\begin{thebibliography}{10}
\providecommand{\url}[1]{#1}
\csname url@samestyle\endcsname
\providecommand{\newblock}{\relax}
\providecommand{\bibinfo}[2]{#2}
\providecommand{\BIBentrySTDinterwordspacing}{\spaceskip=0pt\relax}
\providecommand{\BIBentryALTinterwordstretchfactor}{4}
\providecommand{\BIBentryALTinterwordspacing}{\spaceskip=\fontdimen2\font plus
\BIBentryALTinterwordstretchfactor\fontdimen3\font minus
  \fontdimen4\font\relax}
\providecommand{\BIBforeignlanguage}[2]{{%
\expandafter\ifx\csname l@#1\endcsname\relax
\typeout{** WARNING: IEEEtran.bst: No hyphenation pattern has been}%
\typeout{** loaded for the language `#1'. Using the pattern for}%
\typeout{** the default language instead.}%
\else
\language=\csname l@#1\endcsname
\fi
#2}}
\providecommand{\BIBdecl}{\relax}
\BIBdecl

\bibitem{Berger96}
T.~Berger, Z.~Zhang, and H.~Viswanathan, ``The {CEO} problem,'' \emph{IEEE
  Transactions on Information Theory}, vol.~42, no.~3, pp. 887--902, 1996.

\bibitem{oha98}
Y.~Oohama, ``The rate-distortion function for the quadratic {G}aussian {CEO}
  problem,'' \emph{IEEE Transactions on Information Theory}, vol.~44, no.~3,
  pp. 1057--1070, 1998.

\bibitem{VB97}
H.~Viswanathan and T.~Berger, ``The quadratic {G}aussian {CEO} problem,''
  \emph{IEEE Transactions on Information Theory}, vol.~43, no.~5, pp.
  1549--1559, 1997.

\bibitem{PTR04}
V.~Prabhakaran, D.~Tse, and K.~Ramachandran, ``Rate region of the quadratic
  {G}aussian {CEO} problem,'' in \emph{IEEE International Symposium on
  Information Theory (ISIT), 2004.}\hskip 1em plus 0.5em minus 0.4em\relax
  IEEE, 2004, p. 119.

\bibitem{oha12}
Y.~Oohama, ``Distributed source coding of correlated {G}aussian remote
  sources,'' \emph{IEEE Transactions on Information Theory}, vol.~58, no.~8,
  pp. 5059--5085, 2012.

\bibitem{CZB04}
J.~Chen, X.~Zhang, T.~Berger, and S.~B. Wicker, ``An upper bound on the
  sum-rate distortion function and its corresponding rate allocation schemes
  for the {CEO} problem,'' \emph{IEEE Journal on Selected Areas in
  Communications}, vol.~22, no.~6, pp. 977--987, 2004.

\bibitem{CB08}
J.~Chen and T.~Berger, ``Successive {W}yner-{Z}iv coding scheme and its
  application to the quadratic {G}aussian {CEO} problem,'' \emph{IEEE
  Transactions on Information Theory}, vol.~54, no.~4, pp. 1586--1603, 2008.

\bibitem{BS09}
H.~Behroozi and M.~R. Soleymani, ``Optimal rate allocation in successively
  structured {G}aussian {CEO} problem,'' \emph{IEEE Transactions on Wireless
  Communications}, vol.~8, no.~2, pp. 627--632, 2009.

\bibitem{BS05}
------, ``Performance of the successive coding strategy in the {CEO} problem,''
  in \emph{Global Telecommunications Conference, 2005. GLOBECOM'05. IEEE},
  vol.~3.\hskip 1em plus 0.5em minus 0.4em\relax IEEE, pp. 6--pp.

\bibitem{Tad}
X.~He, X.~Zhou, P.~Komulainen, M.~Juntti, and T.~Matsumoto, ``A lower bound
  analysis of {H}amming distortion for a binary {CEO} problem with joint
  source-channel coding,'' \emph{IEEE Transactions on Communications}, vol.~64,
  no.~1, pp. 343--353, 2016.

\bibitem{ELG}
A.~El~Gamal and Y.-H. Kim, \emph{Network information theory}.\hskip 1em plus
  0.5em minus 0.4em\relax Cambridge university press, 2011.

\bibitem{Tad2}
X.~He, X.~Zhou, M.~Juntti, and T.~Matsumoto, ``A rate-distortion region
  analysis for a binary {CEO} problem,'' in \emph{Vehicular Technology
  Conference (VTC Spring), 2016 IEEE 83rd}.\hskip 1em plus 0.5em minus
  0.4em\relax IEEE, 2016, pp. 1--5.

\bibitem{Tad3}
------, ``Data and error rate bounds for binary data gathering wireless sensor
  networks,'' in \emph{Signal Processing Advances in Wireless Communications
  (SPAWC), 2015 IEEE 16th International Workshop on}.\hskip 1em plus 0.5em
  minus 0.4em\relax IEEE, 2015, pp. 505--509.

\bibitem{RA14}
A.~Razi and A.~Abedi, ``Convergence analysis of iterative decoding for binary
  {CEO} problem,'' \emph{IEEE Transactions on Wireless Communications},
  vol.~13, no.~5, pp. 2944--2954, 2014.

\bibitem{HBP08}
J.~Haghighat, H.~Behroozi, and D.~V. Plant, ``Joint decoding and data fusion in
  wireless sensor networks using turbo codes,'' in \emph{Personal, Indoor and
  Mobile Radio Communications, 2008. PIMRC 2008. IEEE 19th International
  Symposium on}.\hskip 1em plus 0.5em minus 0.4em\relax IEEE, 2008, pp. 1--5.

\bibitem{AW15}
A.~No and T.~Weissman, ``Universality of logarithmic loss in lossy
  compression,'' in \emph{IEEE International Symposium on Information Theory
  (ISIT), 2015}.\hskip 1em plus 0.5em minus 0.4em\relax IEEE, 2015, pp.
  2166--2170.

\bibitem{SRV17}
Y.~Shkel, M.~Raginsky, and S.~Verd{\'u}, ``Universal lossy compression under
  logarithmic loss,'' in \emph{IEEE International Symposium on Information
  Theory (ISIT), 2017}.\hskip 1em plus 0.5em minus 0.4em\relax IEEE, 2017, pp.
  1157--1161.

\bibitem{CW14}
T.~A. Courtade and T.~Weissman, ``Multiterminal source coding under logarithmic
  loss,'' \emph{IEEE Transactions on Information Theory}, vol.~60, no.~1, pp.
  740--761, Jan 2014.

\bibitem{FILL07}
T.~Filler, ``Minimizing embedding impact in steganography using low density
  codes,'' \emph{Master's thesis, SUNY Binghamton}, 2007.

\bibitem{NAA17}
M.~Nangir, M.~Ahmadian, and R.~Asvadi, ``A binary {W}yner-{Z}iv code design
  based on compound {LDGM}-{LDPC} structures,'' \emph{IET Communications, DOI:
  10.1049/iet-com.2017.0032}, 2017.

\bibitem{urban08}
T.~Richardson and R.~Urbanke, \emph{Modern coding theory}.\hskip 1em plus 0.5em
  minus 0.4em\relax Cambridge university press, 2008.

\bibitem{CSV03}
G.~Caire, S.~Shamai, and S.~Verdu, ``Lossless data compression with error
  correcting codes,'' in \emph{Information Theory, 2003. Proceedings. IEEE
  International Symposium on}.\hskip 1em plus 0.5em minus 0.4em\relax IEEE,
  2003, p.~22.

\bibitem{PR03}
S.~S. Pradhan and K.~Ramchandran, ``Distributed source coding using syndromes
  ({DISCUS}): Design and construction,'' \emph{IEEE Transactions on Information
  Theory}, vol.~49, no.~3, pp. 626--643, 2003.

\bibitem{LXG02}
A.~D. Liveris, Z.~Xiong, and C.~N. Georghiades, ``Compression of binary sources
  with side information at the decoder using {LDPC} codes,'' \emph{IEEE
  communications letters}, vol.~6, no.~10, pp. 440--442, 2002.

\bibitem{Remani13}
C.~Remani, ``Numerical methods for solving systems of nonlinear equations,''
  \emph{Lakehead University, Thunder Bay, Ontario, Canada}, 2012-13.

\bibitem{KSA17}
M.~Khas, H.~Saeedi, and R.~Asvadi, ``{LDPC} code design for correlated sources
  using {EXIT} charts,'' in \emph{IEEE International Symposium on Information
  Theory (ISIT), 2017}.\hskip 1em plus 0.5em minus 0.4em\relax IEEE, 2017, pp.
  2945--2949.

\bibitem{SC07}
D.~H. Schonberg, \emph{Practical distributed source coding and its application
  to the compression of encrypted data}.\hskip 1em plus 0.5em minus 0.4em\relax
  University of California, Berkeley, 2007.

\end{thebibliography}
%\bibliography{refs}

\end{document}